\documentclass{jpp}
\usepackage{graphicx,overpic}
\usepackage{epstopdf, epsfig}
\usepackage{hyperref}
\usepackage{color}
\newcommand{\ie}{{i.e.,\/ }}

\shorttitle{The tertiary instability and the Dimits shift within a scalar model}
\shortauthor{H. Zhu, Y. Zhou, and I. Y. Dodin}

\title{Theory of the tertiary instability and the Dimits shift within a scalar model}

\author{Hongxuan Zhu\aff{1,2}
  \corresp{\email{hzhu@pppl.gov}},
  Yao Zhou\aff{1},
 \and I. Y. Dodin\aff{1,2}}

\affiliation{\aff{1}Princeton Plasma Physics Laboratory, Princeton, NJ 08543
\aff{2}Department of Astrophysical Sciences, Princeton University, Princeton,
NJ 08544}

\begin{document}

\maketitle

\begin{abstract}
The Dimits shift is the shift between the threshold of the drift-wave primary instability and the actual onset of turbulent transport in magnetized plasma. It is generally attributed to the suppression of turbulence by zonal flows, but developing a more detailed understanding calls for consideration of specific reduced models. The modified Terry--Horton system has been proposed by St-Onge [J. Plasma Phys. {\bf 83}, 905830504 (2017)] as a minimal model capturing the Dimits shift. Here, we use this model to develop an analytic theory of the Dimits shift and a related theory of the tertiary instability of zonal flows. We show that tertiary modes are localized near extrema of the zonal velocity $U(x)$, where $x$ is the radial coordinate. By approximating $U(x)$ with a parabola, we derive the tertiary-instability growth rate using two different methods and show that the tertiary instability is essentially the primary drift-wave instability modified by the local $U''$. Then, depending on $U''$, the tertiary instability can be suppressed or unleashed. The former corresponds to the case when zonal flows are strong enough to suppress turbulence (Dimits regime), while the latter corresponds to the case when zonal flows are unstable and turbulence develops. This understanding is different from the traditional paradigm that turbulence is controlled by the flow shear $U'$. Our analytic predictions are in agreement with direct numerical simulations of the modified Terry--Horton system.
\end{abstract}

\section{Introduction}

\label{sec:Introduction}

{The Dimits shift in magnetized plasmas is the shift between the threshold of drift-wave (DW) ``primary'' instability and the actual onset of transport that follows the scaling laws of developed turbulence \citep{Dimits00}.}  The Dimits shift is observed in both fluid and gyrokinetic
simulations \citep{Lin98,Rogers00,Ricci06,Numata07,Mikkelsen08,Kobayashi12,St-Onge17} and is generally attributed to turbulence suppression by zonal flows (ZFs), which are generated by the ``secondary'' instability \citep{Rogers00,Diamond01}. However, the Dimits shift is finite, meaning that ZFs cannot completely suppress DW turbulent transport if the primary-instability threshold is exceeded by far. Because of the detrimental effect that turbulent transport has on plasma confinement, it is important to understand this effect in detail.

After the seminal work \citep{Biglari90}, it is widely accepted that ZFs can significantly suppress turbulence by shearing turbulent eddies. Based on this paradigm, the predator--prey
model is perhaps the simplest phenomenological model that can describe
how sheared flows help achieve a high-confinement regime \citep{Diamond94,Malkov01,Kim03,Kobayashi15}. However, this paradigm may be oversimplified. For example, while direct simulations show that ZFs saturate at finite amplitude even in collisionless plasma \citep{Rogers00,St-Onge17}, the predator--prey model predicts otherwise. This is because the predator--prey model assumes statistically homogeneous turbulence, and this assumption is inapplicable in the Dimits regime, where strong ZFs are present and turbulence is inhomogeneous. 

A more elaborate approach to understanding the Dimits shift was based on the concept of the ``tertiary'' instability (TI) \citep{Rogers00,Rogers05}. The idea is that if ZFs are subject to the TI, then turbulence cannot be completely suppressed by ZFs and the Dimits regime ends. Despite some criticism \citep{Kolesnikov05}, this explanation is widely accepted. However, the understanding of the TI and the Dimits shift has been largely qualitative, arguably because these effects have not been widely studied within simple enough models.

Recently, \cite{St-Onge17} proposed the modified Terry--Horton equation (mTHE) as a minimal model that captures the Dimits shift.  \citeauthor{St-Onge17} calculated the TI growth rate using four-mode truncation (4MT) and derived a sufficient condition for ZFs to be stable within the mTHE. Then, this criterion was used for a ``heuristic calculation'' of the Dimits shift. However, that calculation is not entirely satisfactory, because deriving the actual Dimits shift takes more than a sufficient condition of ZF stability. The direct relation between \citeauthor{St-Onge17}'s criterion and the Dimits shift is only an assumption.  As a result, the agreement of \citeauthor{St-Onge17}'s theory with numerical simulations is limited (section~\ref{sec:DSPrediction}). Besides, the 4MT model is only a rough approximation and cannot capture essential features of the TI in principle, as we shall discuss below. Therefore, a transparent theory of the TI and the Dimits shift within the mTHE model is yet to be developed.

In our recent letter \citep{Zhu20}, we sketched a theory of the TI and the Dimits shift within the modified Hasegawa--Wakatani model, where the mTHE was briefly mentioned as the ``adiabatic limit''. This limit is important in that the mTHE permits a detailed analytic study of the TI and an explicit quantitative prediction of the Dimits shift;  thus, it deserves further investigation. Here, we present an in-depth study of the mTHE by expanding on the results presented in \cite{Zhu20}. We show that assuming a sufficient scale separation between  ZFs and  DWs, TI modes are localized at extrema of the ZF velocity $U(x)$, where $x$ is the radial coordinate. By approximating $U(x)$ with a parabola, we analytically derive the TI growth rate, $\gamma_{\rm TI}$, using two different approaches: (i) by drawing an analogy between TI modes and quantum harmonic oscillators and (ii)  by using the Wigner--Moyal equation (WME). Our theory shows that the TI is essentially a primary DW instability modified by the ZF ``curvature'' $U''$ near extrema of $U$. (The prime denotes ${\rm d}/{\rm d}x$.) In particular, the WME helps understand how the local $U''$ modifies the mode structure and reduces the TI growth rate; it also shows that the TI is \textit{not} the Kelvin--Helmholtz (KH) instability, or KHI. Then, depending on $U''$, the TI can be suppressed, in which case ZFs are strong enough to suppress turbulence (Dimits regime), or unleashed, so ZFs are unstable and turbulence develops. This understanding is different from the traditional paradigm \citep{Biglari90}, where turbulence is controlled by the flow shear $U'$.  Finally, by letting $\gamma_{\rm TI}=0$, we obtain an analytic prediction of the Dimits shift, which agrees with our numerical simulations of the mTHE.

{Admittedly, our explicit prediction of the Dimits shift is  facilitated by the fact that we use a simple enough model. Understanding of the Dimits shift is already complicated when we study the modified Hasegawa--Wakatani model in \cite{Zhu20},  when we observed the presence of avalanche-like structures, which are not supported by the mTHE. Furthermore, the recent paper by \cite{Ivanov20} shows that avalanches themselves can become intricate when additional physics from finite ion temperature is taken into account. This complicates the problem even further, and more work remains to be done to understand the Dimits shift in the general case. Our paper is intended  as one of the first steps in that direction.}

This paper is organized as follows. In section~\ref{sec:mTHE} we introduce the mTHE. In section~\ref{sec:Linear} we describe the primary, the secondary, and the tertiary instability within the mTHE. In section~\ref{sec:TIcalculation} we analytically derive the TI growth rate using two different approaches mentioned above. In section~\ref{sec:DSPrediction} we derive an analytic prediction of the Dimits shift. Finally, a brief introduction of the WME and phase-space trajectories are presented in Appendices~\ref{appA} and \ref{appB}.

\section{Modified Terry--Horton equation}
\label{sec:mTHE}
The mTHE can be considered as a minimal model that simultaneously captures the primary, secondary, and tertiary instabilities. It is a two-dimensional scalar equation that describes DW turbulence in slab geometry with coordinates $\boldsymbol{x}=(x,y)$, where $x$ is the radial coordinate and $y$ is the poloidal coordinate: 
\begin{equation}
\label{eq:Equations_mTHE}
\partial_{t}w+\{\varphi,w\}-\beta\partial_{y}\varphi+\hat{\alpha}\hat{D}w=0,
\end{equation}
where
\begin{equation}
w=\nabla^{2}\varphi-n,\quad n=(\hat{\alpha}-{\rm i}\hat{\delta})\varphi.
\end{equation}
Here, the system is assumed to be immersed in a uniform magnetic field
perpendicular to the $(x,y)$ plane. The ions are assumed cold while the
electrons are assumed to have a finite temperature $T_{{\rm e}}$.
The plasma has an equilibrium density profile $n_{0}(x)$, which is parameterized by the positive constant
$\beta\doteq a/L_{n}$, where $a$ is a reference length and $L_{n}\doteq(-{\rm d}\ln n_{0}/{\rm d}x)^{-1}$
is the scale length of the density gradient. (We use $\doteq$ to
denote definitions.) Time is normalized by $a/c_{{\rm s}}$,
where $c_{{\rm s}}\doteq\sqrt{T_{{\rm e}}/m_{{\rm i}}}$ is the ion
sound speed. Length is normalized by the ion sound radius $\rho_{{\rm s}}=c_{{\rm s}}/\Omega_{{\rm i}}$,
where $\Omega_{{\rm i}}$ is the ion gyro-frequency. The electrostatic
potential fluctuation $\varphi$ is normalized by $T_{{\rm e}}\rho_{{\rm s}}/ea$
where $e$ is the unit charge, the electron density fluctuation $n$
is normalized by $n_{0}\rho_{\rm s}/a$, and $w$ can be considered as minus the
ion guiding-center density \citep{Krommes00}. The Poisson bracket is defined as
\begin{equation}
\{\varphi,w\}\doteq\boldsymbol{v}\cdot\nabla w,\quad\boldsymbol{v}\doteq\hat{\boldsymbol{z}}\times\nabla\varphi,\label{eq:Poisson}
\end{equation}
which describes nonlinear advection of $w$ by the $\boldsymbol{E}\times\boldsymbol{B}$
flow with velocity $\boldsymbol{v}$. Also, $\nabla^{2}\doteq\partial_{x}^ {2}+\partial_{y}^{2}$ is the Laplacian. {Finally, we note that the parameter $\beta$ can be scaled out of equation (\ref{eq:Equations_mTHE}) by replacing $(\phi,t,\hat{D})$ with $(\phi/\beta,\beta t,\hat{D}/\beta)$. Therefore, varying $\beta$  is effectively similar to varying the strength of $\hat{D}$}.

The mTHE is ``modified'' compared to the original Terry--Horton model \citep{Terry82,Terry83} in that the following operator $\hat{\alpha}$ is used:
\begin{equation}
\hat{\alpha}\varphi=\tilde{\varphi}\doteq\varphi-\langle\varphi\rangle,\label{eq:Equations_alpha}
\end{equation}
where $\langle\dots\rangle$ is the zonal average  given by 
\begin{equation}
 \langle\varphi\rangle\doteq\frac{1}{L_{y}}\int_{0}^{L_{y}}\varphi\,{\rm d}y
\end{equation}
and $L_{y}$ is the system length along $y$. Equation (\ref{eq:Equations_alpha}) states that
electrons respond only to the fluctuation (or DW) part of the potential, $\tilde{\varphi}$, but do not respond to the zonal-averaged (or ZF) part, $\langle\varphi\rangle$ \citep{St-Onge17,Hammett93}.
The operator $\hat{\delta}$ describes the phase difference between $n$ and $\varphi$ and determines the primary DW instability
\citep{Terry82,Terry83}. Note that (\ref{eq:Equations_mTHE})
reduces to the modified Hasegawa--Mima equation at $\hat{\delta}=0$ \citep{Hasegawa77,Dewar07frontiers}, where the total energy is conserved. The DW and the ZF part of the energy (per unit area) are given by 
\begin{equation}
    E_{\rm DW}\doteq\frac{1}{2L_xL_y}\int{\rm d}x\,{\rm d}y\left[(\nabla_{\perp}\tilde{\varphi})^2+\tilde{
    \varphi}^2\right],\quad E_{\rm ZF}\doteq \frac{1}{2L_x}\int{\rm d}x\,\left(\partial_x\langle\varphi\rangle\right)^2,\label{eq:Energies}
\end{equation}
where $L_x$ is the system length along $x$. Various forms of $\hat{\delta}$ can be used to model different primary instabilities \citep{Terry82,Tang78}. Here,
 we follow \cite{St-Onge17} and use
the following simple form:
\begin{equation}
{\rm i}\hat{\delta}\doteq{\rm i}\delta_{0}\hat{k}_{y}\equiv\delta_{0}\partial_{y},\label{eq:Equations_delta}
\end{equation}
with $\delta_{0}$ being a positive constant. (This can be used to
model trapped-electron dynamics \citep{Tang78}.) Finally, the operator
$\hat{D}$ models damping effects such as viscosity.
 Following \cite{St-Onge17}, we use 
\begin{equation}
\hat{D}=1-0.01\nabla^{2}\label{eq:Equations_D}
\end{equation}
throughout this paper. (An exception is made in section \ref{sec:DSPrediction}, where another form of $\hat{D}$ is  introduced for comparison.) {Here, the first (friction) term  is added in order to prevent possible energy build up at large scale, as is also done by \citeauthor{St-Onge17}. (As will be seen from our results below, this term also increases the Dimits shift and thus facilitates its numerical observation.)} Note that due to  $\hat{\alpha}$ in front
of $\hat{D}$ in (\ref{eq:Equations_mTHE}), the damping applies only to DWs, while ZFs are left collisionless. Then, the Dimits  regime can be defined unambiguously as the regime where ZFs persist forever and the DW amplitude decreases to zero at $t \to \infty$. 

Beyond the Dimits regime, DWs are not suppressed and ZFs always keep evolving in the mTHE model, as demonstrated by \cite{St-Onge17}. To understand the ZF dynamics, we take
the zonal average of (\ref{eq:Equations_mTHE}) and obtain
\begin{equation}
\partial_{t}U=-\partial_{x}\langle\tilde{v}_{x}\tilde{v}_{y}\rangle-\langle\tilde{v}_{x}{\rm i}\hat{\delta}\tilde{\varphi}\rangle+\mathcal{T}(t),\quad U(x,t)\doteq\partial_{x}\langle\varphi\rangle.\label{eq:Equations_Ut}
\end{equation}
Here, $U$ is the ZF velocity along $y$, $(\tilde{v}_{x},\tilde{v}_{y})\doteq(-\partial_{y}\tilde{\varphi},\partial_{x}\tilde{\varphi})$
is the $\boldsymbol{E}\times\boldsymbol{B}$ velocity of  DW fluctuations. The first term on the right-hand side of (\ref{eq:Equations_Ut})
is the Reynolds stress, while the second term is specific to the mTHE
system. For the form of $\hat{\delta}$ given by  (\ref{eq:Equations_delta}),
the second term becomes
\begin{equation}
-\langle\tilde{v}_{x}{\rm i}\hat{\delta}\tilde{\varphi}\rangle=\delta_{0}^{2}\langle\tilde{v}_{x}^{2}\rangle>0.\label{eq:Equation_T(t)}
\end{equation}
Therefore, the second term will always increase the local ZF velocity $U$, and meanwhile, the value of $U$ at other locations will be adjusted by the effect of $\mathcal{T}(t)$, which is an integration constant that ensures conservation of the total momentum. Specifically, $\partial_{t}\int U{\rm d}x=0$ implies
\begin{equation}
\quad\mathcal{T}(t)=\frac{1}{L_x}\int\langle\tilde{v}_{x}{\rm i}\hat{\delta}\tilde{\varphi}\rangle{\rm d}x.
\end{equation}
Due to nonzero $\mathcal{T}$, ZFs cannot remain (quasi)stationary in the presence of fluctuations within the mTHE. In other words, either ZFs completely
suppress  DW turbulence, or both ZFs and DWs keep evolving
indefinitely.

\section{Primary, secondary, and tertiary instability}
\label{sec:Linear}

\begin{figure}
\includegraphics[width=1\columnwidth]{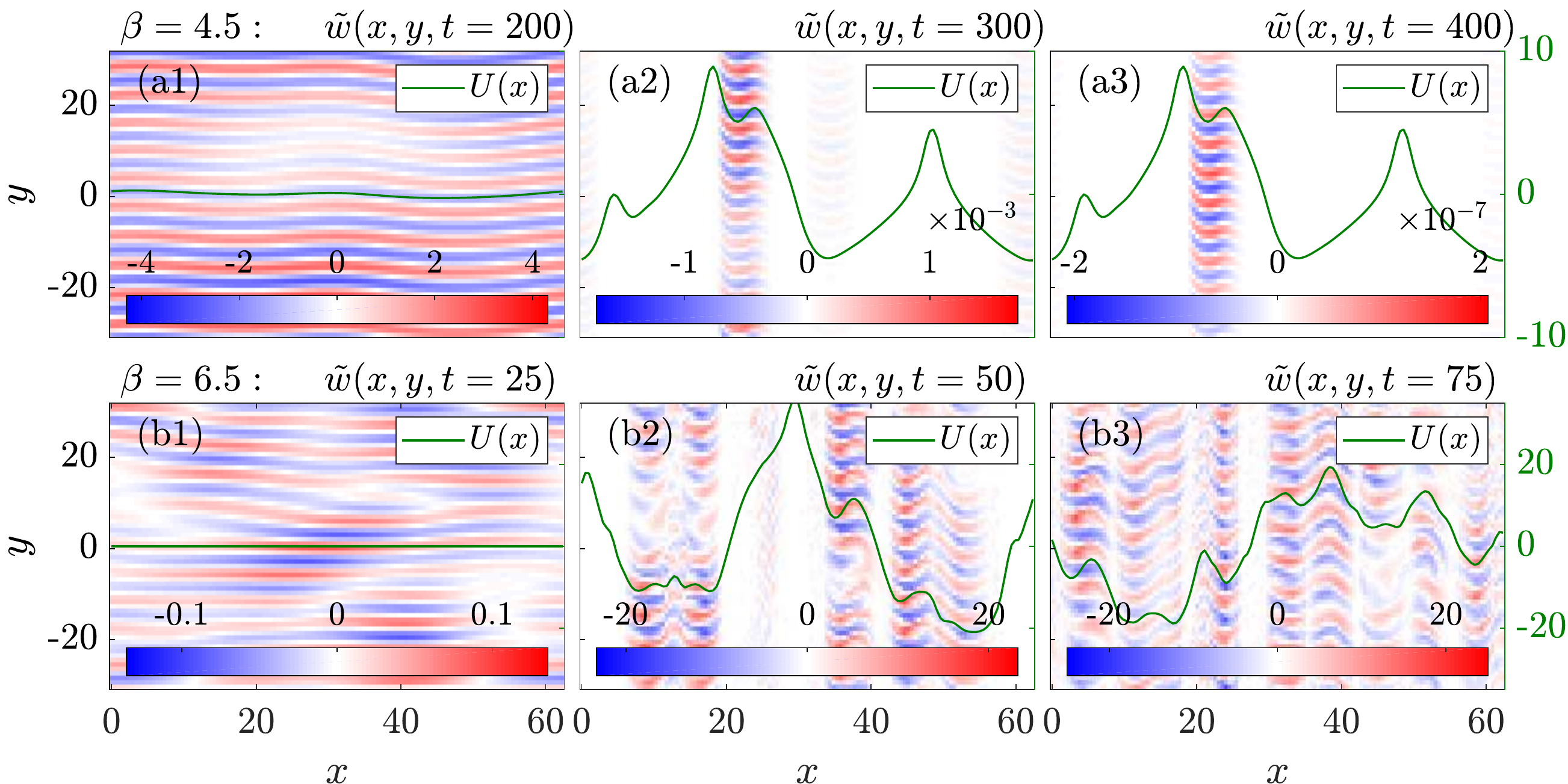}
\caption{Snapshots from numerical
simulations of the mTHE (\ref{eq:Equations_mTHE}) with $\delta_0=1.5$ (see (\ref{eq:Equations_delta})) at  (a) $\beta=4.5$ (first row) and (b) $\beta=6.5$ (second row). The simulation domain size is $L_{x}=L_{y}=20\upi$, with the corresponding numbers of grid points being  $N_x=128$ and $N_y=64$, respectively. Periodic boundary conditions are used in both directions, and the nonlinear term is treated using the pseudospectral method with 2/3  dealiasing rule \citep{boyd2001chebyshev}. The initial conditions are random noise with a small amplitude.  Shown are the fluctuations $\tilde{w}$ (colorbar) and the ZF velocity $U$ (green curve) at three different moments of time. It is seen that at $\beta=4.5$, the DW amplitude decreases down to zero (Dimits regime), while at $\beta=6.5$, fluctuations remain strong and ZFs keep evolving.} 
\label{fig:THhistory}
\end{figure}

We have integrated the mTHE numerically using random noise for the initial conditions. Typical simulation results are presented in figures~\ref{fig:THhistory} and \ref{fig:PI_and_SI}. It is seen that the primary instability of DWs arises and is followed by ZF generation through the secondary instability. Then, at the fully nonlinear stage,  DW turbulence becomes inhomogeneous, exhibiting signatures of the TI. In the following, we study these stages in detail.

\subsection{Primary instability}
It is straightforward to show that $\lbrace \varphi, w \rbrace = 0$  for Fourier eigenmodes of the form
\begin{equation}
\varphi=\varphi_{\boldsymbol{k}}{\rm e}^{{\rm i}\boldsymbol{k}\cdot\boldsymbol{x}-{\rm i}\Omega_{\boldsymbol{k}}t}+{\rm c.c.},
\end{equation}
where $\boldsymbol{k}=(k_{x},k_{y})$. Therefore, a Fourier eigenmode
is an exact solution of the system 
provided that $\Omega_{\boldsymbol{k}}$ satisfies the following relation: 
\begin{equation}
\label{eq:Linear_Primary_dispersion}
\omega_{\boldsymbol{k}}\doteq\Real\,\Omega_{\boldsymbol{k}}=\frac{\beta k_{y}(1+k^{2})}{(1+k^{2})^{2}+\delta_{0}^{2}k_{y}^{2}},\quad
\gamma_{\boldsymbol{k}}\doteq\Imag\,\Omega_{\boldsymbol{k}}=\frac{\beta\delta_{0}k_{y}^{2}}{(1+k^{2})^{2}+\delta_{0}^{2}k_{y}^{2}}-\alpha_{\boldsymbol{k}}D_{\boldsymbol{k}}.\label{eq:Linear_Primary_Omegak}
\end{equation}
Here, $k^{2}\doteq k_{x}^{2}+k_{y}^{2}$, $D_{\boldsymbol{k}}=1+0.01k^{2}$, and we have used (\ref{eq:Equations_delta}).
Also, $\alpha_{\boldsymbol{k}}=1$ for $k_{y}\neq0$ and $\alpha_{\boldsymbol{k}}=0$
for $k_{y}=0$, and hence a ZF ($k_{y}=0$) corresponds to $\Omega_{\boldsymbol{k}}=0$,
\ie to a stationary state. From (\ref{eq:Linear_Primary_Omegak}),
it is seen that when $D_{\boldsymbol{k}}=0$, $\gamma_{\boldsymbol{k}}$
is maximized at $(k_{x},k_{y})=(0,1)$. A nonzero $D_{\boldsymbol{k}}$
can modify the value of $\boldsymbol{k}$ that maximizes $\gamma_{\boldsymbol{k}}$,
but for the chosen form of $\hat{D}$, (\ref{eq:Equations_D}), this modification is very small. Therefore, if one numerically simulates (\ref{eq:Equations_mTHE}) with small random noise as the initial
conditions, then nonlinear interactions can be neglected at first
and coherent DW structures will grow exponentially with typical wavenumber
$\boldsymbol{k}\approx(0,1)$, as seen in figure~\ref{fig:THhistory}.

\subsection{Secondary instability}
\label{subsec:Linear_Secondary}

When many Fourier modes are present and have grown to a finite
amplitude, the nonlinear term in (\ref{eq:Equations_mTHE})  becomes important. This can be seen from the Fourier representation, $\varphi=\sum_{\boldsymbol{k}}\varphi_{\boldsymbol{k}}(t)\exp({\rm i}\boldsymbol{k}\cdot\boldsymbol{x})$,
where (\ref{eq:Equations_mTHE}) is written as
\begin{equation}
\frac{{\rm d}\varphi_{\boldsymbol{k}}}{{\rm d}t}=-{\rm i}\Omega_{\boldsymbol{k}}\varphi_{\boldsymbol{k}}+\frac{1}{2}\sum_{\boldsymbol{k}_{1},\boldsymbol{k}_{2}}T(\boldsymbol{k},\boldsymbol{k}_{1},\boldsymbol{k}_{2})\delta_{\boldsymbol{k},\boldsymbol{k}_{1}+\boldsymbol{k}_{2}}\varphi_{\boldsymbol{k}_{1}}\varphi_{\boldsymbol{k}_{2}}\label{eq:Linear_Primary_Fourier}
\end{equation}
and $\delta_{\boldsymbol{k}_1, \boldsymbol{k}_2}$ is the Kronecker symbol. Also,
\begin{equation}
T(\boldsymbol{k},\boldsymbol{k}_{1},\boldsymbol{k}_{2})\doteq-\frac{\bar{k}_{1}^{2}-\bar{k}_{2}^{2}}{\bar{k}^{2}}(\boldsymbol{k}_{1}\times\boldsymbol{k}_{2})\cdot\hat{\boldsymbol{z}}
\end{equation}
are the coefficients that govern the nonlinear mode coupling, $\bar{k}^{2}$ is defined as
\begin{equation}
\bar{k}^{2}\doteq\alpha_{\boldsymbol{k}}+k^{2}-{\rm i}\delta_{0}k_{y},\label{eq:Equations_kbar}
\end{equation}
and similarly for $\bar{k}_{1}^{2}$ and $\bar{k}_{2}^{2}$.

Due to nonlinear interactions, ZFs can be generated  from DWs, which process is known as the secondary instability. Here, we use the 4MT model to analyze this instability,
namely, by considering a primary DW with  $\boldsymbol{k}=(0,k_{y})$, a
ZF with $\boldsymbol{q}=(q_{x},0)$, and two DW sidebands with $\boldsymbol{k}_{\pm}=(\pm q_{x},k_{y})$.
Assume that the ZF is small, so the exponential
growth of the primary DW is unaffected; \ie  $\varphi_{\boldsymbol{k}}=\varphi_{0}\exp(-{\rm i}\Omega_{\boldsymbol{k}}t$), 
with $\varphi_{0}$ being a constant. Then, from (\ref{eq:Linear_Primary_Fourier}), the
equations that describe the ZF and the sidebands are as follows \citep{St-Onge17}:
\begin{eqnarray}
\label{eq:Linear_Secondary_4MT}
{\rm d}_{t}\varphi_{\boldsymbol{q}}=\frac{k_{y}{\rm e}^{\gamma_{\boldsymbol{k}} t}}{q_{x}}\left[\left(q_{x}^{2}-{\rm i}\delta_{+}\right)\varphi_{\boldsymbol{k}_{+}}\varphi_{0}^{*}{\rm e}^{{\rm i}\omega_{\boldsymbol{k}}t}-\left(q_{x}^{2}+{\rm i}\delta_{+}\right)\varphi_{\boldsymbol{k}_{-}}^{*}\varphi_{0}{\rm e}^{-{\rm i}\omega_{\boldsymbol{k}}t}\right],\label{eq:Linear_Secondary_4MT-a}\\
{\rm d}_{t}\varphi_{\boldsymbol{k}_{+}}=-{\rm i}\Omega_{\boldsymbol{k}_{+}}\varphi_{\boldsymbol{k}_{+}}+T(\boldsymbol{k}_{+},\boldsymbol{k},\boldsymbol{q})\varphi_{0}\varphi_{\boldsymbol{q}}{\rm e}^{-{\rm i}\Omega_{\boldsymbol{k}}t},\\
{\rm d}_{t}\varphi_{\boldsymbol{k}_{-}}=-{\rm i}\Omega_{\boldsymbol{k}_{-}}\varphi_{\boldsymbol{k}_{-}}+T(\boldsymbol{k}_{-},\boldsymbol{k},-\boldsymbol{q})\varphi_{0}\varphi_{\boldsymbol{q}}^{*}{\rm e}^{-{\rm i}\Omega_{\boldsymbol{k}}t},
\end{eqnarray}
where $\delta_{+}\doteq\delta_{\boldsymbol{k}}+\delta_{\boldsymbol{k}_{+}}=2\delta_{0}k_{y}$. We have also used $\Omega_{\boldsymbol{k}}=\omega_{\boldsymbol{k}}+{\rm i}\gamma_{\boldsymbol{k}}$.
These equations can be combined to yield a single time-evolution equation
for the ZF amplitude $\varphi_{\boldsymbol{q}}$:
\begin{equation}
\frac{{\rm d}^{3}\varphi_{\boldsymbol{q}}}{{\rm d}t^{3}}-A\,\frac{{\rm d}^{2}\varphi_{\boldsymbol{q}}}{{\rm d}t^{2}}+(B-C)\,\frac{{\rm d}\varphi_{\boldsymbol{q}}}{{\rm d}t}-D\varphi_{\boldsymbol{q}}=0.\label{eq:Linear_Secondary_phiqt}
\end{equation}
Here, $A=2\gamma_{+}$,  $B=\omega_{-}^{2}+\gamma_{+}^{2}$,  $C,D\propto|\varphi{}_{0}{\rm e}^{\gamma_{\boldsymbol{k}}t}|^{2}$, $\gamma_{+}\doteq\gamma_{\boldsymbol{k}}+\gamma_{\boldsymbol{k}_{+}}$, and $\omega_{-}\doteq\omega_{\boldsymbol{k}}-\omega_{\boldsymbol{k}_{+}}$.
The derivation of (\ref{eq:Linear_Secondary_phiqt}) can be found
in \citeauthor{St-Onge17}. Expressions for $C$ and $D$ can also be found there but will not be important for our discussion; however, note that compared to \citeauthor{St-Onge17}, we have absorbed the coefficient ${\rm e}^{\gamma_{\boldsymbol{k}}t}$ into the definitions of $C$ and $D$.

When $C$ and $D$ are much larger than $A$ and $B$, $\varphi_{\boldsymbol{q}}$
can grow ``super-exponentially'' \citep{Rogers00,St-Onge17}, \ie as an exponential of an exponential. This is also known
as the secondary KH instability \citep{Rogers00}. In the opposite case, when $A$ and $B$ dominate over $C$ and $D$, the non-constant solution of (\ref{eq:Linear_Secondary_phiqt}) is approximately
\begin{equation}
\varphi_{\boldsymbol{q}}\propto{\rm e}^{(\gamma_{+}\pm{\rm i}\omega_{-})t}.
\end{equation}
Since $\gamma_{+}$
decreases as $|q_{x}|$ increases (see (\ref{eq:Linear_Primary_Omegak})), the growth rate is maximized at
the lowest ZF wavenumber $|q_{x}|=2\upi/L_{x}$. In other words,
the box-scale ZF grows fastest, with the growth rate given by $\gamma_{+}\approx2\gamma_{\boldsymbol{k}}$, i.e., twice the growth rate of the primary DW instability. 

\begin{figure}
\includegraphics[width=1\columnwidth]{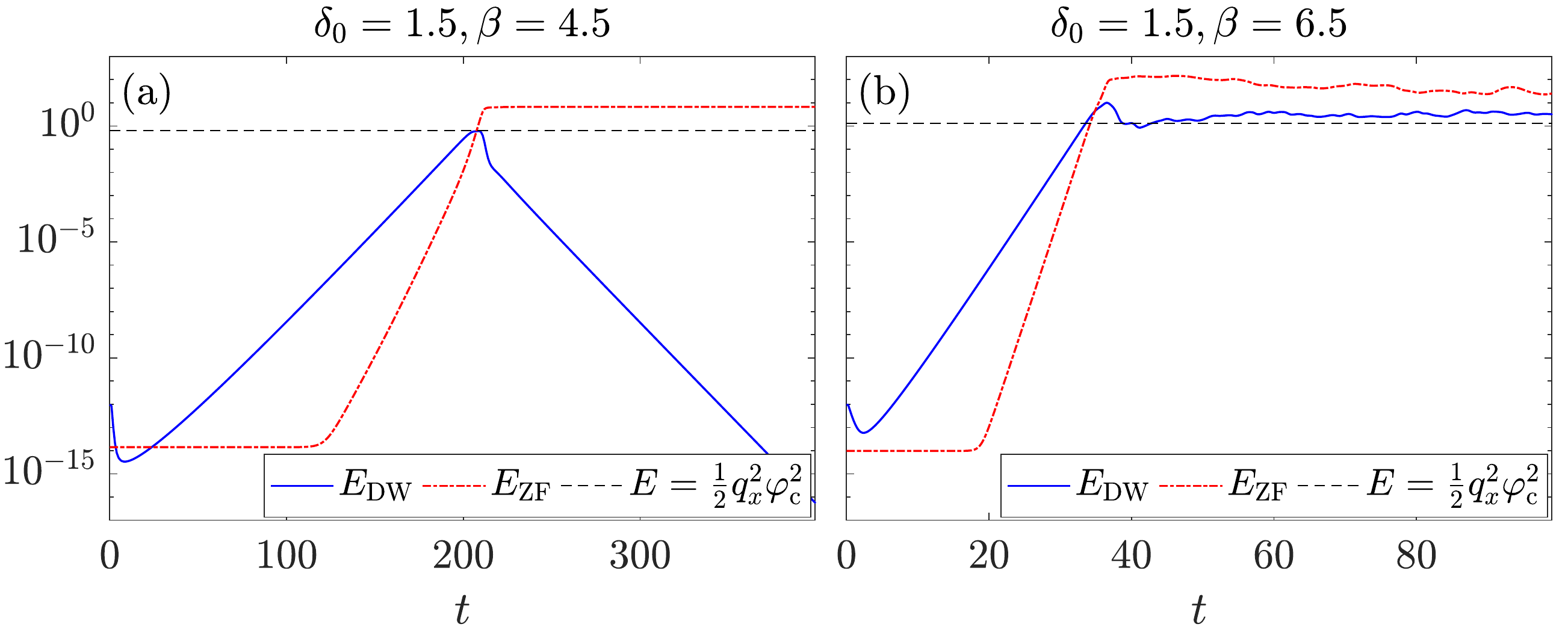}
\caption{The time history of the DW and ZF energies (\ref{eq:Energies}) corresponding to figure~\ref{fig:THhistory}. The primary and secondary instabilities are clearly seen, with the 
secondary-instability growth rate being twice the primary-instability growth rate.  The
black dashed line is the ZF energy calculated from (\ref{eq:Linear_Secondary_Ec})
that corresponds to the critical ZF amplitude $\varphi_{{\rm c}}$
from (\ref{eq:Linear_Secondary_phic}). It is seen that this energy roughly corresponds to the onset of the fully nonlinear regime. This value is reached by both $E_{\rm DW}$ and $E_{\rm ZF}$  at approximately the same time.}
\label{fig:PI_and_SI}
\end{figure}

In the following, we
show that exponential growth of the ZF at the box scale is more common than the super-exponential growth, provided that the characteristic amplitude $\varphi_0$ of the initial random noise is small enough.
At first, both the primary DW and the sidebands grow exponentially,
\begin{equation}
|\varphi_{\boldsymbol{k}}|\sim|\varphi_{\boldsymbol{k}_{\pm}}|\sim|\varphi_{{\rm 0}}|{\rm e}^{\gamma_{\boldsymbol{k}}t},
\end{equation}
while the ZF amplitude remains at the noise level. Then, DWs grow for some time $t_{{\rm p}}$ before they
begin to affect ZFs. Assume that at $t=t_{{\rm p}}$,  the box-scale
ZF with the amplitude $\varphi_{\boldsymbol{q}}\sim\varphi_{0}$ starts to grow with the growth rate $\gamma_{+}\approx2\gamma_{\boldsymbol{k}}$;
then, $\delta_{+}=2\delta_{0}k_{y}\gg q_{x}^{2}$, and we have from (\ref{eq:Linear_Secondary_4MT-a}) that
\begin{equation}
    |\partial_{t}\varphi_{\boldsymbol{q}}|\sim 2\gamma_{\boldsymbol{k}}|\varphi_{{\rm 0}}|\approx\frac{2|\varphi_{\boldsymbol{k}}||\varphi_{\boldsymbol{k}_{+}}|k_{y}\delta_{+}}{q_{x}}\approx\frac{2k_{y}\delta_{+}|\varphi_{{\rm 0}}{\rm e}^{\gamma_{\boldsymbol{k}}t_{{\rm p}}}|^{2}}{q_{x}}.\label{eq:Linear_Secondary_phi0}
\end{equation}
This leads to
\begin{equation}
     C,D\propto\frac{q_{x}\gamma_{\boldsymbol{k}}|\varphi_{0}|}{2\delta_{0}k_{y}^{2}}.
\end{equation}
Therefore, $C$ and $D$ are small when the initial noise level $|\varphi_{0}|$
is small enough; hence, the assumptions made above are self-consistent, namely, $A$ and $B$ are indeed much larger than $C$ and $D$, and
the box-scale ZF with wavenumber $q_{x}=2\pi/L_{x}$ grows fastest with the growth rate $2\gamma_{\boldsymbol{k}}$.

The secondary instability will persist for some
time $t_{{\rm s}}$ until ZFs grows up to a finite amplitude that is enough to 
significantly distort the DW structure. Using the result from  \cite{Zhu18b}, this amplitude can be estimated as follows (also see (\ref{eq:appB_U_critical})):
\begin{equation}
\varphi_{{\rm c}}=\frac{\beta/q_{x}}{2(1+k_{y}^{2})-q_{x}^{2}}.\label{eq:Linear_Secondary_phic}
\end{equation}
At $\varphi_{\boldsymbol{q}}\ll\varphi_{{\rm c}}$, DWs do not
``see'' the ZF and hence keep growing exponentially, while at $\varphi_{\boldsymbol{q}}\gtrsim\varphi_{{\rm c}}$
the system enters the fully
nonlinear regime. Therefore, $t_{{\rm s}}$ is the time when the ZF
amplitude grows from $\varphi_{0}$ to $\varphi_{{\rm c}}$, and it can be estimated as follows:
\begin{equation}
t_{{\rm s}}=\frac{1}{2\gamma_{\boldsymbol{k}}}\ln\frac{\varphi_{{\rm c}}}{\varphi_{0}}.\label{eq:ts}
\end{equation}
Note that (\ref{eq:Linear_Secondary_phic}) is obtained from the modified Hasegawa--Mima system, so it is based on the assumption that $\delta_0 = 0$. For nonzero $\delta_0$, it is modified accordingly (see (\ref{eq:appB_U_critical})), but the above estimate is sufficient for our qualitative description. 

By the time when the system enters the fully nonlinear regime, the DW amplitude becomes $|\varphi_{\boldsymbol{k}}|\sim\varphi_0\exp{\gamma_{\boldsymbol{k}}(t_{\rm s}+t_{\rm p})}$, which can be estimated from (\ref{eq:Linear_Secondary_phi0}) and (\ref{eq:ts}) as
\begin{equation}
|\varphi_{\boldsymbol{k}}|\sim\sqrt{\frac{q_{x}\gamma_{\boldsymbol{k}}\varphi_{\rm c}}{2\delta_{0}k_{y}^{2}}}.
\end{equation}
From (\ref{eq:Energies}), the corresponding DW and ZF energies are as follows:
\begin{equation}
E_{{\rm ZF}}\sim\frac{\beta^{2}}{8(1+k_{y}^{2})^{2}},\quad
E_{{\rm DW}}\sim\frac{\beta\gamma_{\boldsymbol{k}}}{8\delta_{0}k_{y}^{2}},\label{eq:Linear_Secondary_Ec}
\end{equation}
where we assumed $q_x^2\ll 1+k_y^2$. Using (\ref{eq:Linear_Primary_Omegak}) for $\gamma_{\boldsymbol{k}}$ and assuming $D_{\boldsymbol{k}}=0$ for simplicity, we obtain
\begin{equation}
\frac{E_{{\rm ZF}}}{E_{{\rm DW}}}\sim 1 + \frac{\delta_0^2k_y^2}{(1+k_y^2)^2}\,.
\end{equation}
This shows that the ZF energy and the DW energy are roughly equal to each other when the system enters the fully nonlinear regime, since $\delta_0$ and $k_y$ are of order unity. This conclusion
will be used to estimate the ZF curvature in section~\ref{sec:DSPrediction}.

These predictions are in agreement with numerical simulations (figure~\ref{fig:PI_and_SI}). This indicates that the 4MT captures the basic dynamics of the primary and the secondary instabilities. However, as shown below, the 4MT does not capture essential features of the TI, and thus more accurate models are needed to describe the TI and the Dimits shift.

\subsection{Tertiary instability}

\label{subsec:Linear_Tertiary}

\begin{figure}
\includegraphics[width=1\textwidth]{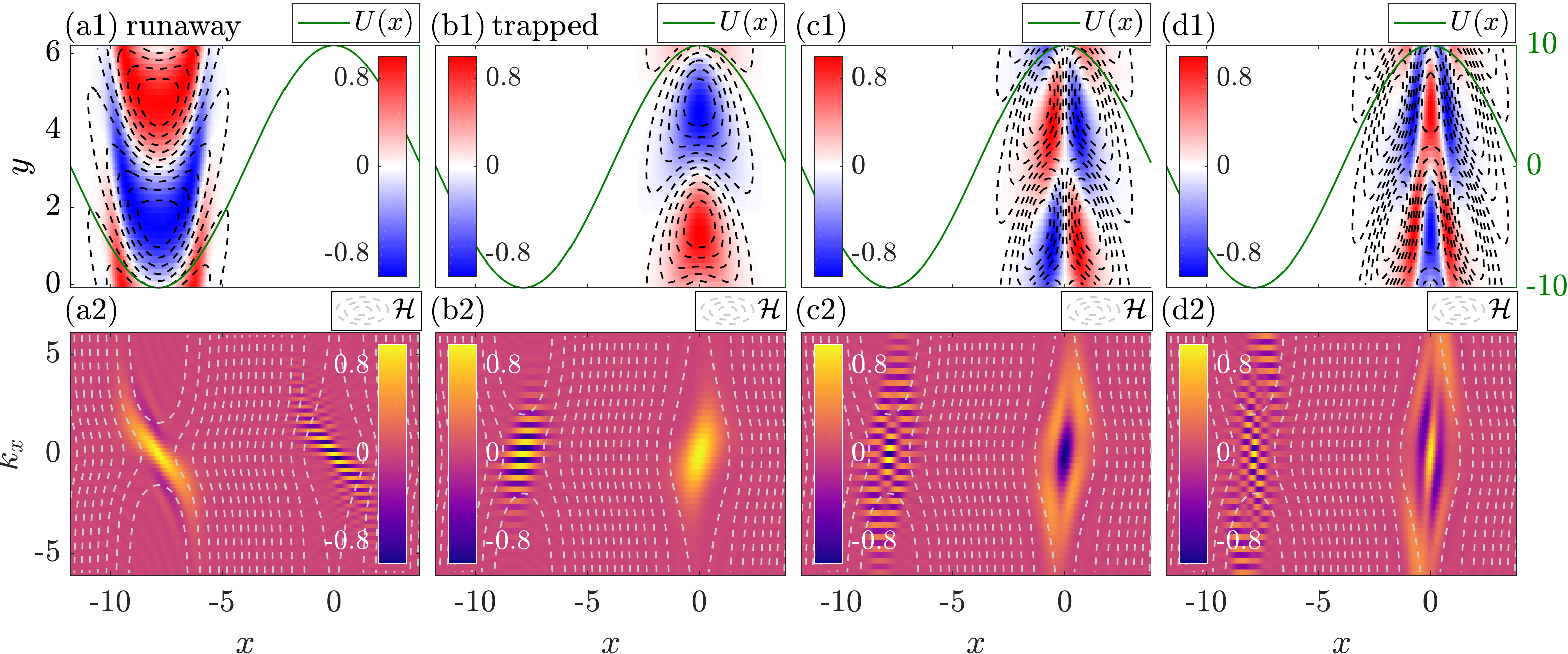}
\caption{The first four tertiary eigenmodes found numerically using the ZF velocity profile (\ref{eq:Linear_Tertiary_sinusoidalU}). The ordering is such that $\gamma_{\rm TI}$ decreases from left to right.  The first two eigenmodes are runaway and trapped modes, respectively. The parameters
are $\beta=6$, $\delta_{0}=1.5$, $q_{x}=0.4$, and $u=10$. The first row shows the eigenmode structures $\tilde{w}(x,y)=\Real[w(x){\rm e}^{{\rm i}k_{y}y}]$ ((\ref{eq:Linear_Tertiary_eigenmode}), color), the ZF velocity $U$ (green curve), and the analytic mode structure $\tilde{w}=\Real[\mathsf{H}_m(x)\exp(S+{\rm i}k_y y)]$ ((\ref{eq:AnalyticTI_S}), dashed contour), where $m=0$ for (a1) and (b1), $m=1$ for (c1), and $m=2$ for (d1). The second row shows the corresponding Wigner function $W(x,k_x)$ ((\ref{eq:appA_Wigner}), color) and the isosurfaces of the drifton Hamiltoninan $\mathcal{H}$ ((\ref{eq:appB_Drifton_Hamiltonian}), dashed contour).  The striped structure of $W$ away from
the actual location of DW quanta is a signature of a quantumlike  ``cat state'' \citep{Weinbub18}.} \label{fig:TI_eigenvalue}
\end{figure}

In the fully nonlinear regime, DW turbulence becomes inhomogeneous and localized at the extrema of the ZF velocity $U$ (figure~\ref{fig:THhistory}). To understand the DW dynamics in this case, let us linearize (\ref{eq:Equations_mTHE}) to obtain
\begin{equation}
\partial_{t}\tilde{w}+U\partial_{y}\tilde{w}-(\beta+U'')\partial_{y}\tilde{\varphi}+\hat{D}\tilde{w}=0,\label{eq:Linear_Tertiary_linearDW}
\end{equation}
where
\begin{equation}
\tilde{w}=(\nabla^{2}-1+{\rm i}\hat{\delta})\tilde{\varphi},\quad U''\doteq{\rm d}^{2}U(x)/{\rm d}x^{2}.
\end{equation}
For given boundary conditions in $x$, eigenmodes of (\ref{eq:Linear_Tertiary_linearDW}) can be searched for in the form
\begin{equation}
\tilde{w}=w(x){\rm e}^{{\rm i}(k_{y}y-\omega t)},\quad\tilde{\varphi}=\left(\frac{{\rm d^{2}}}{{\rm d}x^{2}}-k_{y}^{2}-1+{\rm i}\delta_{0}k_{y}\right)^{-1}\tilde{w},\label{eq:Linear_Tertiary_eigenmode}
\end{equation}
which leads to the following equation for $w(x)$:
\begin{equation}
\omega w=\hat{H}w,\quad \hat{H}(\hat{x},\hat{k}_x)\doteq k_{y}\hat{U}+k_{y}(\beta+\hat{U}'')\hat{\bar{k}}^{-2}-{\rm i}\hat{D},\label{eq:Linear_Tertiary_eigen}
\end{equation}
where
\begin{equation}
\hat{U}=U(\hat{x}),\quad \hat{k}_x=-{\rm i}\,{\rm d}/{\rm d}x,\quad\hat{\bar{k}}^{2}=1+k_{y}^{2}+\hat{k}_x^2-{\rm i}\delta_{0}k_{y}.\label{eq:Linear_Tertiary_khat}
\end{equation}
If an eigenvalue $\omega$
exists and
\begin{equation}
\gamma_{{\rm TI}}\doteq\Imag\,\omega>0,
\end{equation}
then the perturbation grows exponentially. This is the TI. 

Equation (\ref{eq:Linear_Tertiary_linearDW}) does not have an analytic solution for an arbitrary profile $U$, but a general understanding can be developed by considering special cases. 
In \cite{Zhu18c}, we considered the ZF velocity profile
\begin{equation}
U(x)=u\cos q_{x}x,\label{eq:Linear_Tertiary_sinusoidalU}
\end{equation}
with $\hat{\delta} = \hat{D} = 0$. In this case, the system exhibits an instability of the KH type provided that
$q_{x}^{2}>1$ and $q_x^2u>\beta$. In \cite{Zhu18c}, we also discussed a generalization to periodic nonsinusoidal profiles. However, generalizing those results to nonzero $\hat{\delta}$ and $\hat{D}$ is challenging. The common approach is to adopt the 4MT again, \ie to assume a DW perturbation with $\boldsymbol{k}=(0,k_{y})$
and two sidebands with $\boldsymbol{k}_{\pm}=(\pm q_{x},k_{y})$ as small
perturbations \citep{Kim02,St-Onge17,Rath18,Zhu18a}. In particular,
\cite{St-Onge17} derived $\gamma_{{\rm TI}}$
within the 4MT and estimated the Dimits shift by finding a sufficient condition for $\gamma_{{\rm TI}}=0$. However, the 4MT-based approach
is not entirely satisfactory, because the ZF is typically far from sinusoidal, as seen in simulations. Even more importantly, the 4MT approach ignores the fact that there are multiple TI modes with different growth rates. As we show below, understanding the variety of these modes is essential for understanding the Dimits shift.

Let us assume the same sinusoidal ZF profile (\ref{eq:Linear_Tertiary_sinusoidalU}) as in \citeauthor{St-Onge17} for now, and let us calculate the corresponding eigenmodes (\ref{eq:Linear_Tertiary_eigen}) numerically, assuming periodic boundary conditions $x$. In this case, we can search for solutions in the form
\begin{equation}
w(x)=\sum_{n=-N}^{N}w_{n}{\rm e}^{{\rm i}nq_{x}x},
\end{equation}
where $N$ is some large enough integer. {In other words, we truncate the Fourier series by keeping only the first $2N+1$ Fourier modes.} This turns (\ref{eq:Linear_Tertiary_eigen}) into a vector equation for $\{w_{-N}, \dots w_0, \dots w_N\}$, where $\hat{H}$ becomes a $(2N+1)\times(2N+1)$ matrix. Then, one finds $2N+1$ eigenmodes with complex eigenfrequencies. Typical numerical eigenmodes are illustrated in figure~\ref{fig:TI_eigenvalue}.  It is seen that the TI-mode structure is localized at the maximum
($x=0$) or minimum ($x=-\upi/q_{x}$) of the ZF velocity and has either even or odd parity
because of the symmetry of $U$. Within the figure, the eigenmodes localized at the ZF minimum can be labeled by the integer
$m=0,1,2,\dots$, which also indicates the parity of $w(x)$. Eigenmodes localized near the ZF maximimum can be labeled similarly. Note that in order for a mode to be localized, the ZF must be large-scale, namely, $q_x^2\ll 1+k_y^2$, which is consistent with numerical simulations.

Apart from the eigenmode structures, we also show in figure~\ref{fig:TI_eigenvalue}
their corresponding Wigner functions $W(x,k_x)$  (\ref{eq:appA_Wigner}) and contour plots of the drifton Hamiltonian $\mathcal{H}$
(\ref{eq:appB_Drifton_Hamiltonian}). The Wigner function
can be understood as the distribution function of  ``driftons''
(DW quanta) in the $(x,k_{x})$ phase space \citep{Smolyakov99,Ruiz16,Zhu18c}, and its shape is expected to align with the contours of $\mathcal{H}$ . Then, eigenmodes are naturally centered at phase-space equilibria of $\mathcal{H}$, namely,
\begin{equation}
\partial_{x}\mathcal{H}=\partial_{k_{x}}\mathcal{H}=0\quad\Rightarrow\quad U' = k_x=0.\label{eq:Linear_Tertiary_localize}
\end{equation}
This explains eigenmode localization near extrema of $U$. {[Strictly speaking, (\ref{eq:Linear_Tertiary_localize}) stems from our approximation of sinusoidal flow (\ref{eq:Linear_Tertiary_sinusoidalU}), which ensures that $U'$ and $U'''$ become zero at same locations. Nevertheless, (\ref{eq:Linear_Tertiary_localize}) remains a good approximation as long as ZFs are large-scale, \ie $|U'''/\bar{k}^2|\ll U'$}.] Maxima of $U$ (even $n$) correspond to phase-space islands encircled by ``trapped'' trajectories, and minima of $U$ (odd $n$) correspond to saddle points passed by the
``runaway'' trajectories \citep{Zhu18a,Zhu18b,Zhu18c}.  Hence, we call the modes localized near maxima and minima of $U$ trapped and runaway modes, respectively. (See Appendix~\ref{appB} for more discussions on drifton phase-space trajectories.) In the next section, we provide
analytic calculation of the TI growth rates based on the above observations.

\section{Tertiary-instability growth rate}
\label{sec:TIcalculation}
\subsection{Analogy with a quantum harmonic oscillator}
\label{subsec:Analytic_oscillator}
As seen in figure~\ref{fig:TI_eigenvalue}, tertiary modes are centered at the phase-space equilibria. Based on this, let us expand the Hamiltonian up to the second order both in $x$ and in $\hat{k}_x$. Specifically,  we approximate the ZF velocity with a parabola:
\begin{equation}
     U\approx U_0+\frac{1}{2}\,\mathcal{C}x^2,\label{eq:AnalyticTI_oscillator_Uexpand}
\end{equation} 
where $U_0$ is the local ZF velocity and $\mathcal{C}\doteq U''(0)$ is the local ZF curvature. For the sinusoidal velocity (\ref{eq:Linear_Tertiary_sinusoidalU}), this corresponds to $U_0=\pm u$ and $\mathcal{C}=\mp q_x^2 u$. We also make the approximation that $\hat{D}\approx D_0\doteq D_{\boldsymbol{k}=(0,k_y)}$ and
\begin{equation}
   \quad \hat{k}^{-2}\approx k_0^{-2}+k_0^{-4}\frac{\rm d^2}{{\rm d}x^2},\quad k_0^2\doteq 1+k_y^2-{\rm i}\delta_0 k_y.\label{eq:AnalyticTI_oscillator_Kexpand}
\end{equation}
 Then, the Hamiltonian operator $\hat{H}$ (\ref{eq:Linear_Tertiary_eigen}) is approximated as
\begin{equation}
    \hat{H}\approx k_y U_0 +\frac{1}{2}k_y\mathcal{C}\hat{x}^2+k_y(\beta+\mathcal{C})\left(k_0^{-2}+k_0^{-4}\frac{\rm d^2}{{\rm d}x^2}\right)-{\rm i}D_0,\label{eq:AnalyticTI_oscillator_Hexpand}
\end{equation}
and the corresponding eigenmode equation  (\ref{eq:Linear_Tertiary_eigen}) becomes
\begin{equation}
\left(-\tau^2\frac{\rm d^2}{{\rm d}x^2}+x^2\right)w=\varepsilon w.\label{eq:AnalyticTI_oscillator}
\end{equation}
It is the same equation that describes a quantum harmonic oscillator, except that here the coefficients are complex; specifically,
\begin{equation}
    \tau^2\doteq-\frac{2}{k_0^4}\left(1+\frac{\beta}{\mathcal{C}}\right),\quad\varepsilon\doteq\frac{2[\omega-k_y U_0+{\rm i}D_0-k_y(\beta+\mathcal{C})/k_0^2]}{k_y\mathcal{C}}.\label{eq:AnalyticTI_oscillator_coefficients}
\end{equation}
Note that the coefficients are different at minima and maxima of $U$, as they depend on the sign of $\mathcal{C}$. Also note that for runaway modes, we have shifted the coordinate as $x\to x+\upi/q_{x}$ to recenter the ZF minimum at $x=0$.

Following the standard procedure known from quantum mechanics \citep{Sakurai94book}, one can show that the asymptotic behavior of the solution at large $|x|$ is 
\begin{equation}
    w(x)\sim {\rm e}^{S(x)},\quad S(x)=-\frac{x^2}{2\tau}=-\left[\frac{{\rm i}(1+k_y^2)+\delta_0k_y}{2\sqrt{2(1+\beta/\mathcal{C})}}\right]x^2.\label{eq:AnalyticTI_S}
\end{equation}
To ensure that $w\to 0$ at large $|x|$, we require ${\rm Im}\sqrt{1+\beta/\mathcal{C}}\,>0$ if $1+\beta/\mathcal{C}<0$. We also assumed that $\delta_0,k_y>0$. Then, letting $w=\phi(x)\exp S(x)$, we obtain
\begin{equation}
    \phi''-\frac{2x}{\tau}\phi'+\frac{\varepsilon-\tau}{\tau^2}\phi=0.\label{eq:Hermite}
\end{equation}
Solutions are   $\phi=\mathsf{H}_m(x/\sqrt{\tau})$, where $\mathsf{H}_m$ are  Hermite polynomials, $m=0,1,2,\dots$, and
\begin{equation}
    \varepsilon=(2m+1)\tau.
\end{equation}
Therefore, for each sign of $\mathcal{C}$, eigenmodes are labeled by $m$. In figure~\ref{fig:TI_eigenvalue}, these approximate solutions are compared with numerical solutions of (\ref{eq:Linear_Tertiary_eigen}). In the following, we shall focus on the two modes with $m = 0$, since they are most unstable. In this case, $\phi=\mathsf{H}_0$ is  constant and $\varepsilon=\tau$. This corresponds to $\tilde{w}=\Real[\exp(S+{\rm i}k_y y)]$, and the eigenfrequencies are found from (\ref{eq:AnalyticTI_oscillator_coefficients}) to be
\begin{equation}
    \omega=\bar{\Omega}+k_yU_0-\frac{{\rm i}k_y\mathcal{C}\sqrt{(1+\beta/\mathcal{C})/2}}{1+k_y^2-{\rm i}\delta_0 k_y},\quad \bar{\Omega}=\frac{k_y(\beta+\mathcal{C})}{k_0^2}-{\rm i}D_0.\label{eq:AnalyticTI_oscillator_omega}
\end{equation}
Here, $\bar{\Omega}$ is the primary-mode eigenfrequency $\Omega_{\boldsymbol{k}}$ (\ref{eq:Linear_Primary_Omegak}) modified by $\mathcal{C}$, $k_yU_0$ is the local Doppler shift, and the remaining term in $\omega$ vanishes at zero $\mathcal{C}$. Note that at $\mathcal{C}=0$, $\omega$ reduces to the primary-mode frequency $\Omega_{\boldsymbol{k}}$ at $\boldsymbol{k}=(0,k_y)$.  Hence, TI modes found here can be interpreted as standing primary modes modified by ZFs. Accordingly,  the TI growth rate $\gamma_{\rm TI}$ approaches the primary-instability growth rate in the limit $\mathcal{C} \to 0$.

{Let us examine the validity of our approximation in (\ref{eq:AnalyticTI_oscillator_Hexpand}). First, the parabolic approximation of $U$ is valid if the mode spatial width in $x$, which is determined by $\mathcal{C}$, is much smaller than $q_x^{-1}$, which is the characteristic scale of ZFs. Specifically, for the sinusoidal ZF (\ref{eq:Linear_Tertiary_sinusoidalU}), we have $\mathcal{C}=q_x^2 u$, so  the parabolic approximation (\ref{eq:AnalyticTI_oscillator_Uexpand}) is valid at small enough $q_x$ and large enough $u=\mathcal{C}/q_x^2$. Second, the expansion of $\hat{k}^{-2}$  in  (\ref{eq:AnalyticTI_oscillator_Kexpand}) is valid at $|k_x^2|\ll |k_0^2|$, where $k_x$ is the characteristic mode wavenumber in $x$. From (\ref{eq:AnalyticTI_oscillator}), $k_x$ can be estimated as $k_x=1/\sqrt{\tau}$.  Then, the requirement $|k_x^2|\ll |k_0^2|$ leads to $|\sqrt{2(1+\beta/\mathcal{C})}|\gg 1$, or equivalently, $|\mathcal{C}|\ll\beta$, and one expects that the approximation in (\ref{eq:AnalyticTI_oscillator_Kexpand}) becomes invalid as $|\mathcal{C}|$ approaches $\beta$. Therefore, in the following, we restrict our consideration to the parameter regime  $|\mathcal{C}|<\beta$, which is also the regime relevant to our numerical simulations.}

The TI growth rate $\gamma_{\rm TI}$ is obtained by taking the imaginary part of $\omega$.  Within the regime $|\mathcal{C}|<\beta$, let us introduce the notation
\begin{equation}
    \bar{\gamma}\doteq{\rm Im}\,\bar{\Omega}=\frac{\delta_0k_y^2(\beta+\mathcal{C})}{(1+k_y^2)^2+\delta_0^2k_y^2}-D_0,
\end{equation}
which is the primary-instability growth rate $\gamma_{\boldsymbol{k}}$ (\ref{eq:Linear_Primary_Omegak}) modified by $\mathcal{C}$. Then, for the runaway mode (labeled with superscript ``R''), which corresponds to $\mathcal{C}>0$, one has
\begin{equation}
    \gamma_{\rm TI}^{\rm R}=\bar{\gamma}-\sqrt{\frac{1}{2}\left(1+\frac{\beta}{\mathcal{C}}\right)}\,\frac{(1+k_y^2)k_y\mathcal{C}}{(1+k_y^2)^2+\delta_0^2k_y^2}. \label{eq:Analytic_gammaR}
\end{equation}
For the trapped mode (labeled with superscript ``T''), which corresponds to $\mathcal{C}<0$, one has
\begin{equation}
    \gamma_{\rm TI}^{\rm T}=\bar{\gamma}+\sqrt{\frac{1}{2}\left(\left|\frac{\beta}{\mathcal{C}}\right|-1\right)}\,\frac{\delta_0k_y^2\mathcal{C}}{(1+k_y^2)^2+\delta_0^2k_y^2}.\label{eq:Analytic_gammaT}
\end{equation}
Figure~\ref{fig:TIgrowthRate}(c)  shows that these formulas are in good agreement with our numerical calculations of the eigenvalues. {Also, we have verified (not shown) that the results in figure~\ref{fig:TIgrowthRate}(c) are insensitive to $q_x$ as long as $q_x$ is small, more specifically, $q_x^2\ll 1+k_y^2$.}

Notably, while the trapped-mode growth rate always decreases with $|\mathcal{C}|$, the runaway-mode growth rate can increase at large $\mathcal{C}$ if $\delta_0$ is large. In fact, at $\mathcal{C}\gg\beta$, (\ref{eq:Analytic_gammaR}) becomes
\begin{equation}
    \gamma_{\rm TI}^{\rm R}\approx \left[\frac{\delta_0k_y^2}{(1+k_y^2)^2+\delta_0^2k_y^2}-\sqrt{\frac{1}{2}}\,\frac{(1+k_y^2)k_y}{(1+k_y^2)^2+\delta_0^2k_y^2}\right]\mathcal{C},
\end{equation}
which predicts that $\gamma_{\rm TI}^{\rm R}$ increases with $\mathcal{C}$ if {$\delta_0>(k_y + k_y^{-1})/\sqrt{2}$. } Therefore, it is possible that the TI can develop in strong ZFs, but the physical mechanism is very different from the KH mode, as will be discussed in section~\ref{subsec:KHI}.  (Strictly speaking, (\ref{eq:Analytic_gammaR}) becomes invalid at $\mathcal{C}\gg\beta$. Nevertheless, we have verified from numerical calculations (not shown) that at $k_y=1$, $\gamma_{\rm TI}^{\rm R}$ indeed increases with $\mathcal{C}$ at large $\mathcal{C}$, if $\delta_0\gtrsim 1.7$.) 

\subsection{Alternative approach}
\label{subsec:Analytic_WME}
\begin{figure}
\includegraphics[width=1\columnwidth]{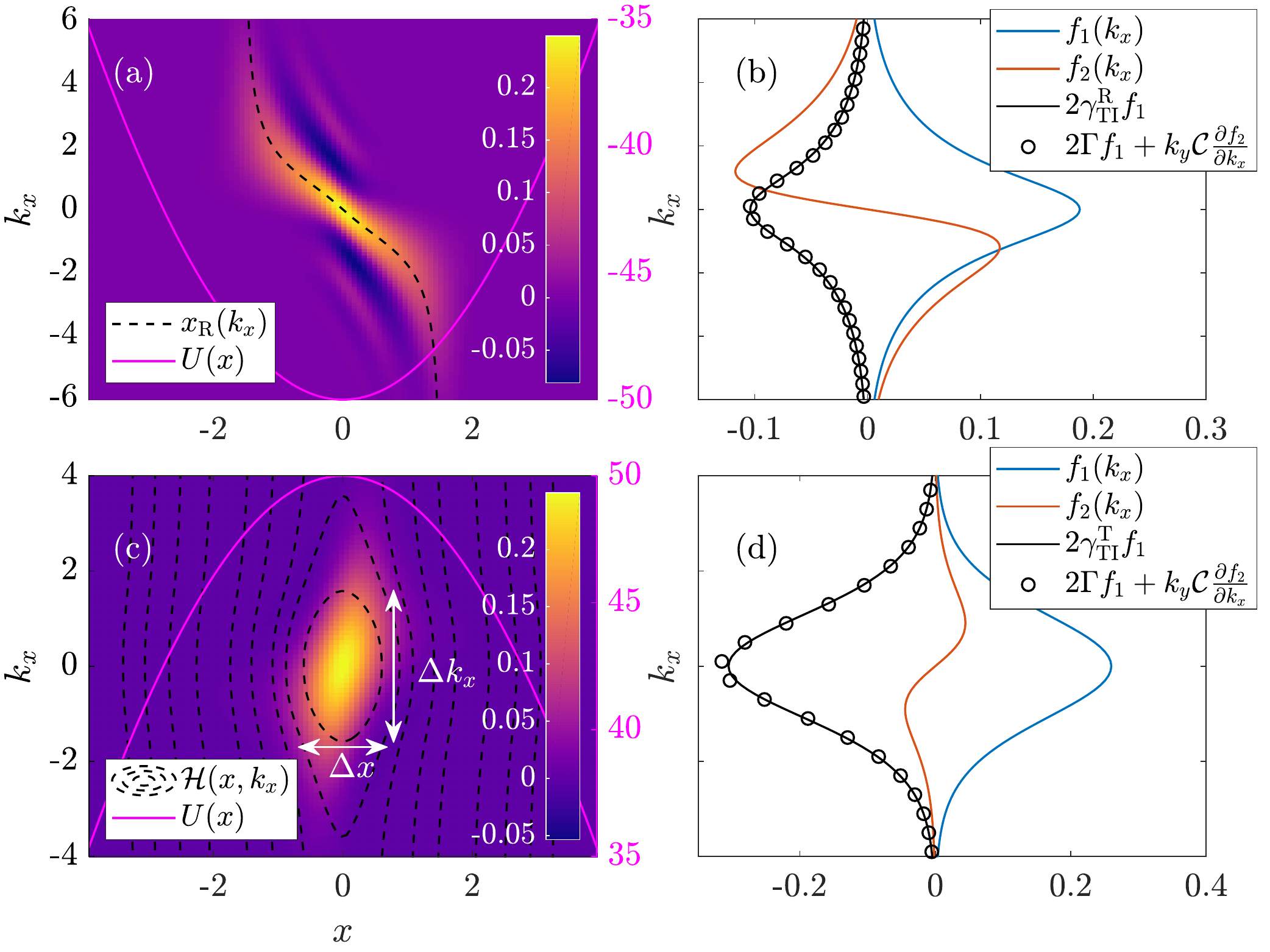}
\caption{The structure of the tertiary modes with $m=0$ in the zonal-velocity profile  (\ref{eq:Linear_Tertiary_sinusoidalU}). The parameters are $\delta_{0}=1.6$, $\beta=5$, $q_{x}=0.2$,
$u=50$ (hence, $\mathcal{C}=\pm 2$), and $\hat{D}$ given by (\ref{eq:Equations_D}). These parameters result in $\gamma_{{\rm TI}}^{{\rm R}}=-0.276$ and $\gamma_{{\rm TI}}^{{\rm T}}=-0.587$. (a) The Wigner function $W(x,k_x)$ of the runaway mode (color), the local $U$ (magenta curve), and the runaway trajectory (dashed curve; see (\ref{eq:GammaTI_runaway_traj})). {Note that we have shifted the coordinates as $x\to x+\pi/q_x$ to recenter the ZF minimum at $x=0$.} (b) The structure of each term in (\ref{eq:GammaTI_xintegration}) calculated from $W$ of the runaway mode in (a). (c) The Wigner function $W(x,k_x)$ of the trapped mode (color), the local $U$ (magenta curve), and isosurfaces of $\mathcal{H}$ (dashed contours; see (\ref{eq:GammaTI_Hamiltonian})). In this figure, $\Delta x$ and $\Delta k_x$ denote the characteristic widths of the mode in the $x$ and $k_x$ directions, correspondingly. (d) Same as (b) but for the trapped mode.  \label{fig:TIstructure}}
\end{figure}

\begin{figure}
\includegraphics[width=1\columnwidth]{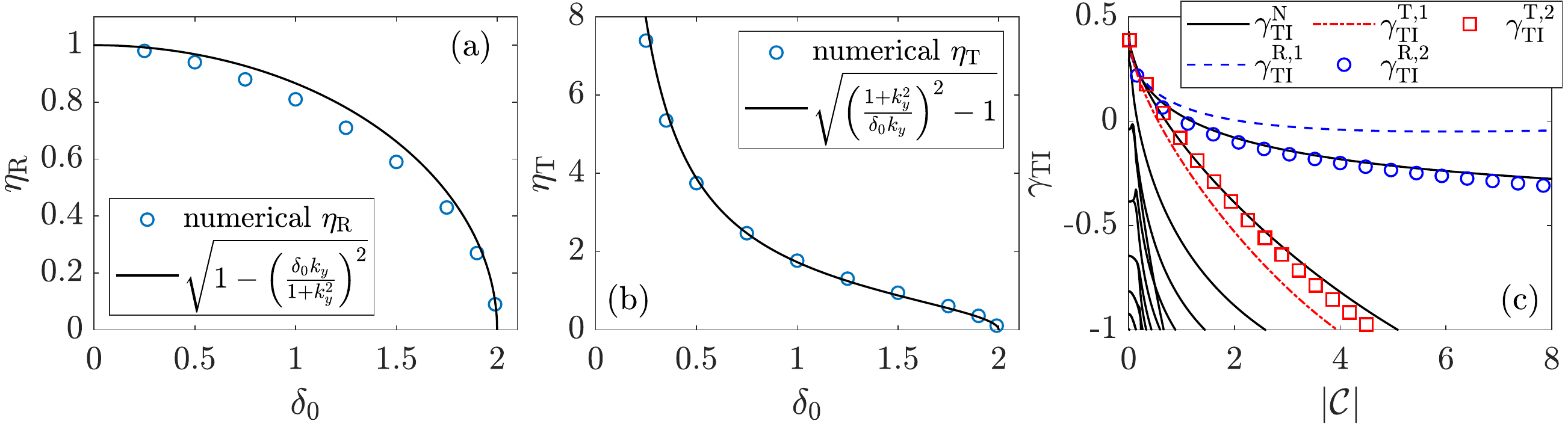}
\caption{(a) The empirical factor $\eta_{\rm R}$ (\ref{eq:GammaTI_runaway_eta}) as a function for $\delta_0$: numerical values {of a sinusoidal ZF (\ref{eq:Linear_Tertiary_sinusoidalU})} (blue circles) versus the fitting formula (black curve). The parameters are $\beta=6$, $q_x=0.4$, $u=10$, and $\hat{D}$ given by (\ref{eq:Equations_D}). It is found that $\eta_{\rm R}$ is not sensitive to $u$. (b) Same as (a) except for $\eta_{\rm T}$ (\ref{eq:GammaTI_trapped_eta}).  It is found that $\eta_{\rm T}$ is not sensitive to $u$ at $u< \beta/q_x^2$. 
(c) The TI growth rates versus $|\mathcal{C}|=q_x^{2}u$ at $\delta_{0}=1.5$, $\beta=6$, $q_{x}=0.4$, and varying $u$. Black curves: numerical solutions of (\ref{eq:Linear_Tertiary_eigen}) indicated by the superscript ``N''. Multiple branches are shown, with the two most unstable branches being the runaway mode and the trapped mode. Blue dashed curve and red dash-dotted curve:  analytic formulas (\ref{eq:Analytic_gammaR}) and (\ref{eq:Analytic_gammaT}). The superscript ``1'' corresponds to predictions made using the approach described in section~\ref{subsec:Analytic_oscillator}. Blue circles and red squares: analytic formulas  (\ref{eq:GammaTI_runaway_GrowthRate}) with $\eta_{{\rm R}}=0.595$ and (\ref{eq:GammaTI_trapped_GrowthRate})
with $\eta_{{\rm T}}=1.2$. The superscript ``2'' corresponds to predictions made using the approach described in section~\ref{subsec:Analytic_WME}.} \label{fig:TIgrowthRate}
\end{figure}

An alternative formula for $\gamma_{\rm TI}$ can be obtained using the Wigner--Moyal equation (WME) for the Wigner function $W$ of the fluctuations $\tilde{w}$ (Appendix \ref{appA}). This approach is somewhat more accurate because the Hamiltonian is expanded only in $x$ but not in $k_x$. As in section~\ref{subsec:Analytic_oscillator}, let us assume $U=U_0+\mathcal{C}x^2/2$. Then, $U''=\mathcal{C}$ is constant, $U'''$ vanishes, and the drifton Hamiltonian 
is simplified down to (Appendix~\ref{appA})
\begin{equation}
    \mathcal{H}=k_{y}U_{0}+\frac{1}{2}k_{y}\mathcal{C}x^{2}+k_{y}(\beta+\mathcal{C})\Real\left(\frac{1}{\bar{k}^{2}}\right),\quad\Gamma=k_{y}(\beta+\mathcal{C}){\rm Im}\left(\frac{1}{\bar{k}^{2}}\right)-D_{\boldsymbol{k}},\label{eq:GammaTI_Hamiltonian}
\end{equation}
where $\bar{k}^2=1+k_x^2+k_y^2-{\rm i}\delta_0k_y$.  Then, the WME (\ref{eq:Drifton_WignerMoyal}) acquires the form
\begin{equation}
\frac{\partial W}{\partial t} = k_{y}\mathcal{C}x\,\frac{\partial W}{\partial k_{x}}-V_{{\rm g}}\,\frac{\partial W}{\partial x}+2\Gamma W+\frac{\partial Q}{\partial x},\label{eq:GammaTI_WME}
\end{equation}
where
\begin{equation}
V_{{\rm g}}(k_{x})\doteq\frac{\partial\mathcal{H}}{\partial k_{x}}=k_{y}(\beta+\mathcal{C})\frac{\partial\Real(1/\bar{k}^{2})}{\partial k_{x}}
\end{equation}
is the drifton group velocity. (Details of drifton dynamics are discussed in Appendix~\ref{appB}.) The value of $Q$ is given by (\ref{eq:Drifton_Q}), but it is not important for our calculations, because we are interested only in the spatial integral of  (\ref{eq:GammaTI_WME}). Since $V_{{\rm g}}$
and $\Gamma$ are independent of $x$, integrating (\ref{eq:GammaTI_WME}) over $x$ leads to
\begin{equation}
2\gamma_{{\rm TI}}f_{1}=k_{y}\mathcal{C}\,\frac{\partial f_{2}}{\partial k_{x}}+2\Gamma f_{1},\label{eq:GammaTI_xintegration}
\end{equation}
where we have replaced $\partial_{t}$ with $2\gamma_{{\rm TI}}$ and introduced
\begin{equation}
     f_{1}(k_{x})\doteq\int W{\rm d}x,\quad f_{2}(k_{x})\doteq\int xW{\rm d}x.
\end{equation}
The functions are shown in figure~\ref{fig:TIstructure} for the runaway mode and for the trapped mode, respectively. {Note that from comparing (\ref{eq:GammaTI_xintegration}) with (\ref{eq:AnalyticTI_oscillator_omega}), it is seen that $f_1$ is associated with the modified frequency $\bar{\Omega}$, namely, $\Gamma=\bar{\gamma}=\Imag\bar{\Omega}$; meanwhile, $\partial f_2/\partial k_x$ is associated with the additional term in (\ref{eq:AnalyticTI_oscillator_omega}) that vanishes at $\mathcal{C}=0$.}

To obtain $\gamma_{\rm TI}$ from (\ref{eq:GammaTI_xintegration}), one needs to find the relation between $f_1$ and $f_2$. Let us first consider the runaway mode. As shown in figure~\ref{fig:TIstructure}(a), the Wigner function of this mode peaks along $x=x_{{\rm R}}(k_{x})$, which is the runaway trajectory that passes through {the saddle point of $\mathcal{H}$ at} $x=k_x=0$, and is given by (\ref{eq:GammaTI_runaway_traj}) below. Therefore, let us adopt $f_{2}\approx x_{{\rm R}}f_{1}$; then,
\begin{equation}
\frac{\partial f_{2}}{\partial k_{x}}\approx\frac{\partial x_{{\rm R}}}{\partial k_{x}}f_{1}+x_{{\rm R}}\frac{\partial f_{1}}{\partial k_{x}}.\label{eq:GammaTI_runaway_f2f3}
\end{equation}
With this assumption, let us evaluate (\ref{eq:GammaTI_xintegration}) at
$k_{x}=0$, where $\partial f_{1}/\partial k_{x}=0$ {because $f_1$ is even in $k_x$ due to the symmetry of $\hat{H}$ (see figure \ref{fig:TIstructure})}; then, we find
\begin{equation}
\gamma_{{\rm TI}}^{{\rm R}}=\left(\Gamma+\frac{k_{y}\mathcal{C}}{2\eta_{{\rm R}}}\frac{\partial x_{{\rm R}}}{\partial k_{x}}\right)\bigg|_{k_{x}=0}.\label{eq:GammaTI_runaway_GrowthRate}
\end{equation}
Here, the first term $\Gamma$ is given by (\ref{eq:GammaTI_Hamiltonian}). The second term is negative because $\partial x_{{\rm R}}/\partial k_{x}<0$ (see (\ref{eq:GammaTI_runaway_slope}) below). The coefficient $\eta_{{\rm R}}>0$ is an empirical factor that compensates for the inaccuracy of (\ref{eq:GammaTI_runaway_f2f3}).   We proceed to determine
$x_{{\rm R}}(k_{x})$ and $\eta_{{\rm R}}$. The runaway trajectory
$x_{{\rm R}}$ is determined from (\ref{eq:GammaTI_Hamiltonian})
by equating $\mathcal{H}$ to its value at the origin $(x,k_{x})=(0,0)$ and
solving $x$ as a function of $k_{x}$. This gives
\begin{equation}
x_{{\rm R}}(k_{x})=\pm\sqrt{2\left(1+\frac{\beta}{\mathcal{C}}\right)}
\sqrt{\frac{1+k_{y}^{2}}{(1+k_{y}^{2})^{2}+\delta_{0}^{2}k_{y}^{2}}-\frac{1+k_{y}^{2}+k_{x}^{2}}{(1+k_{y}^{2}+k_{x}^{2})^{2}+\delta_{0}^{2}k_{y}^{2}}},\label{eq:GammaTI_runaway_traj}
\end{equation}
{where the plus sign is for $k_x<0$ and the minus sign is for $k_x>0$.} 
Figure~\ref{fig:TIstructure}(a) demonstrates that this solution  indeed correlates well with the actual runaway-mode structure. Also note that $x_{{\rm R}}$ is finite, namely,
\begin{equation}
x_{{\rm R}}(k_{x}=\infty)=-\sqrt{2\left(1+\frac{\beta}{\mathcal{C}}\right)\frac{1+k_{y}^{2}}{(1+k_{y}^{2})^{2}+\delta_{0}^{2}k_{y}^{2}}}.\label{eq:GammaTI_xR}
\end{equation}
From
 (\ref{eq:GammaTI_runaway_traj}), we obtain
\begin{equation}
\frac{\partial x_{{\rm R}}}{\partial k_{x}}\bigg|_{k_{x}=0}=-\frac{\sqrt{2(1+\beta/\mathcal{C})\left[(1+k_{y}^{2})^{2}-\delta_{0}^{2}k_{y}^{2}\right]}}{(1+k_{y}^{2})^{2}+\delta_{0}^{2}k_{y}^{2}}.\label{eq:GammaTI_runaway_slope}
\end{equation}
Notably, $\partial x_{{\rm R}}/\partial k_{x}$ becomes zero at $\delta_{0}=|k_{y}+k_{y}^{-1}|$, which corresponds to the transition from runaway to trapped trajectory at the ZF minimum,
as shown in figure~\ref{fig:THtraj}.  

Now, let us consider the correction factor $\eta_{\rm R}$, which can be formally defined as
\begin{equation}
\eta_{{\rm R}}\doteq\left(\frac{\partial x_{{\rm R}}}{\partial k_{x}}\,\frac{f_{1}}{\partial f_{2}/\partial k_{x}}\right)\bigg|_{k_{x}=0}.\label{eq:GammaTI_runaway_eta}
\end{equation}
We determine $\eta_{{\rm R}}$ numerically from
the eigenmode structures obtained in section~\ref{subsec:Linear_Tertiary}.
It can be shown that if $\hat{D}=0$, then rescaling {
$t\to {k_y\mathcal{C}t}/{(1+k_y^2)}$, $x\to x\sqrt{1+k_y^2}$, and $k_x\to {k_x}/{\sqrt{1+k_y^2}}$ leaves only two parameters in the WME (\ref{eq:GammaTI_WME}), namely,}
$\mathcal{C}/\beta$ and $\delta_0k_y/(1+k_y^2)$;  hence, $\eta_{{\rm R}}$ mainly depends on these two parameters. Numerically, we see that $\eta_{{\rm R}}$ changes little
 as $\mathcal{C}/\beta$ varies from zero to unity.
Meanwhile, the dependence of $\eta_{{\rm R}}$ on $\delta_{0}k_{y}/(1+k_{y}^{2})$ is shown in figure~\ref{fig:TIgrowthRate}(a), which suggests the following approximation:
\begin{equation}
\eta_{{\rm R}}\approx\sqrt{1-\left(\frac{\delta_{0}k_{y}}{1+k_{y}^{2}}\right)^{2}}\,.\label{eq:GammaTI_runaway_fitting}
\end{equation}
Then, (\ref{eq:GammaTI_runaway_GrowthRate}) is simplified as
\begin{equation}
\gamma_{{\rm TI}}^{{\rm R}}\approx\Gamma|_{k_{x}=0}-\sqrt{\frac{1}{2}\left(1+\frac{\beta}{\mathcal{C}}\right)}\frac{(1+k_{y}^{2})k_{y}\mathcal{C}}{(1+k_{y}^{2})^{2}+\delta_{0}^{2}k_{y}^{2}}.\label{eq:GammaTI_runaway_connection}
\end{equation}
Remarkably, this formula is identical to (\ref{eq:Analytic_gammaR}) that was obtained in section~\ref{subsec:Analytic_oscillator} by drawing the analogy with a quantum harmonic oscillator.

The above approach can also be applied to the trapped mode. Similarly to (\ref{eq:GammaTI_runaway_GrowthRate}), the trapped-mode growth rate can be expressed as follows:
\begin{equation}
\gamma_{{\rm TI}}^{{\rm T}}=\left(\Gamma+\frac{k_{y}\mathcal{C}}{2\eta_{{\rm T}}}\,\frac{\Delta x}{\Delta k_{x}}\right)\bigg|_{k_{x}=0},\label{eq:GammaTI_trapped_GrowthRate}
\end{equation}
where
\begin{equation}
\frac{\Delta x}{\Delta k_{x}}\doteq\frac{\sqrt{2(|\beta/\mathcal{C}|-1)\left[(1+k_{y}^{2})^{2}-\delta_{0}^{2}k_{y}^{2}\right]}}{(1+k_{y}^{2})^{2}+\delta_{0}^{2}k_{y}^{2}},\quad\eta_{{\rm T}}\doteq\left(\frac{\Delta x}{\Delta k_{x}}\,\frac{f_{1}}{\partial f_{2}/\partial k_{x}}\right)\bigg|_{k_{x}=0}.\label{eq:GammaTI_trapped_eta}
\end{equation}
Here, $\mathcal{C}<0$, and we consider the regime $\beta/\mathcal{C}<-1$. Also, $\Delta x/\Delta k_{x}$ is not the slope of the runaway trajectory but the ratio of the $x$-axis
radii and the $k_{x}$-axis radii of the elliptic trapped trajectories
near $(x,k_{x})=(0,0)$ in figure~\ref{fig:TIstructure}(c). ($\Delta x/\Delta k_{x}$ becomes zero at $\delta_0=|k_y+k_y^{-1}|$, which corresponds to the transition from a single island to two islands, as shown in Fig.~\ref{fig:THtraj}.) The coefficient $\eta_{\rm T}$ is determined numerically. As shown in figure~\ref{fig:TIgrowthRate}(b), $\eta_{\rm T}$ can be approximated as
\begin{equation}
    \eta_{\rm T}\approx\sqrt{\left(\frac{1+k_y^2}{\delta_0k_y}\right)^2-1}\label{eq:GammaTI_trapped_fitting}
\end{equation}
at $\beta/\mathcal{C}<-1$, when the mode is well localized in phase space. In this case, (\ref{eq:GammaTI_runaway_GrowthRate}) becomes identical to (\ref{eq:Analytic_gammaT}).

These results show that the alternative approach adopted here is in agreement with the one we used in section~\ref{subsec:Analytic_oscillator} if we use the fitting formula (\ref{eq:GammaTI_runaway_fitting}) for $\eta_{\rm R}$ and (\ref{eq:GammaTI_trapped_fitting}) for $\eta_{\rm T}$. If these factors are calculated numerically instead, then the alternative approach is slightly more accurate, as seen in figure~\ref{fig:TIgrowthRate}(c).

\subsection{Connection with the Kelvin--Helmholtz instability}\label{subsec:KHI}
\begin{figure}
\includegraphics[width=1\columnwidth]{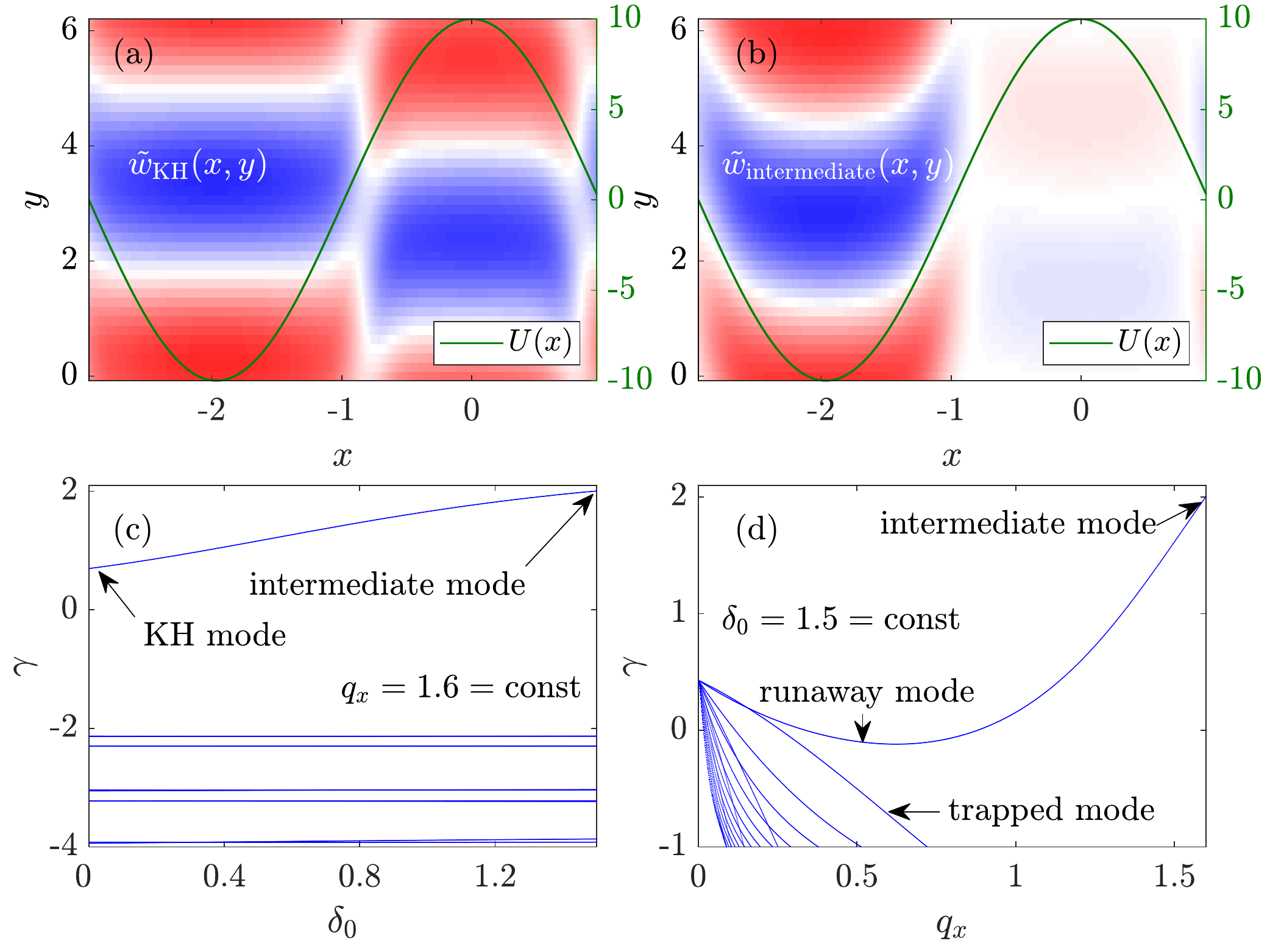}
\caption{Numerical solutions of (\ref{eq:Linear_Tertiary_eigen}) illustrating the relation between the runaway mode and the KH mode at $\beta = 6$, $u = 10$, and various $q_x$ and $\delta_0$. (a) At $\delta_0 = 0$ and $q_x = 1.6$, the unstable mode is the KH mode, which has a global structure as discussed in \cite{Zhu18a,Zhu18c}. (b) The KH mode transitions to an ``intermediate'' mode as $\delta_0$ is increased from $\delta_0=0$ to $\delta_0 = 1.5$ while keeping $q_x = 1.6$ fixed. (c) The corresponding evolution of $\gamma$ with $\delta_0$ at constant $q_x = 1.6$. Blue curves show multiple branches of eigenmodes, but only one branch (the KH mode) is unstable. (d) $\gamma$ as a function of $q_x$ at constant $\delta_0 = 1.5$. As $q_x$ decreases, the intermediate mode analytically continues into the runaway TI mode in figure~\ref{fig:TI_eigenvalue}. See the main text for details.
}\label{fig:KHI}
\end{figure}

The above analysis shows that the TI can be considered as a  primary instability modified by ZFs. As seen from figure~\ref{fig:TIgrowthRate}, the growth rate $\gamma_{\rm TI}$ decreases with $|\mathcal{C}|$ in general. Therefore, the TI is very different from the KHI, which develops only in strong ZFs. To study the relation between the TI and the KHI, we numerically solve (\ref{eq:Linear_Tertiary_eigen}) for various $q_{x}$ and $\delta_{0}$ and explore how the mode structure changes with these parameters. The results are shown in figure~\ref{fig:KHI}. 

First, consider figure~\ref{fig:KHI}(a), which shows a   global (not localized) KH mode that corresponds to  $q_x = 1.6$ and $\delta_0 = 0$. This KH mode has been discussed in \cite{Zhu18a}; it is global because the ZF is small-scale, specifically, $q_x^2 > 1$. Next, let us increase $\delta_0$ from zero up to $\delta_0=1.5$ while keeping $q_x=1.6$ fixed. Then, the original KH mode transforms into an ``intermediate'' mode shown in figure~\ref{fig:KHI}(b). It is not a pure KHI, because dissipation (\ie nonzero $\delta_0$) is now important, but it is not quite the TI either, because $q_x^2$ is large and the mode localization is less pronounced. Our  theory does not apply to such modes, but we have calculated the growth rate numerically as a function of $\delta_0$, as shown in figure~\ref{fig:KHI}(c). Finally, with $\delta_0 = 1.5$ fixed, let us reduce $q_x$. The mode localization improves and the instability rates goes down at first, as seen in figure~\ref{fig:KHI}(d). But eventually, when $q_x$ has become small enough ($q_x \sim 0.6$), the mode transforms into the runaway mode that we introduced earlier (figure~\ref{fig:TIstructure}) and our theory becomes applicable. 

This shows that in principle, the KH mode can be continuously transformed into the runaway mode. However, the KHI and TI are fundamentally different in physical mechanisms, because the TI is due to dissipation and $\gamma_{\rm TI}^{\rm R}$ is determined by $\delta_0$, while the KHI requires a strongly sheared flow and has $\gamma_{\rm KHI} \sim k_y u$. Since typical large-scale ZFs seen in simulations have $q_x^2 \ll 1$, the TI is more relevant to them than the KHI.

\section{Dimits shift}

\label{sec:DSPrediction}
\begin{figure}
\includegraphics[width=1\columnwidth]{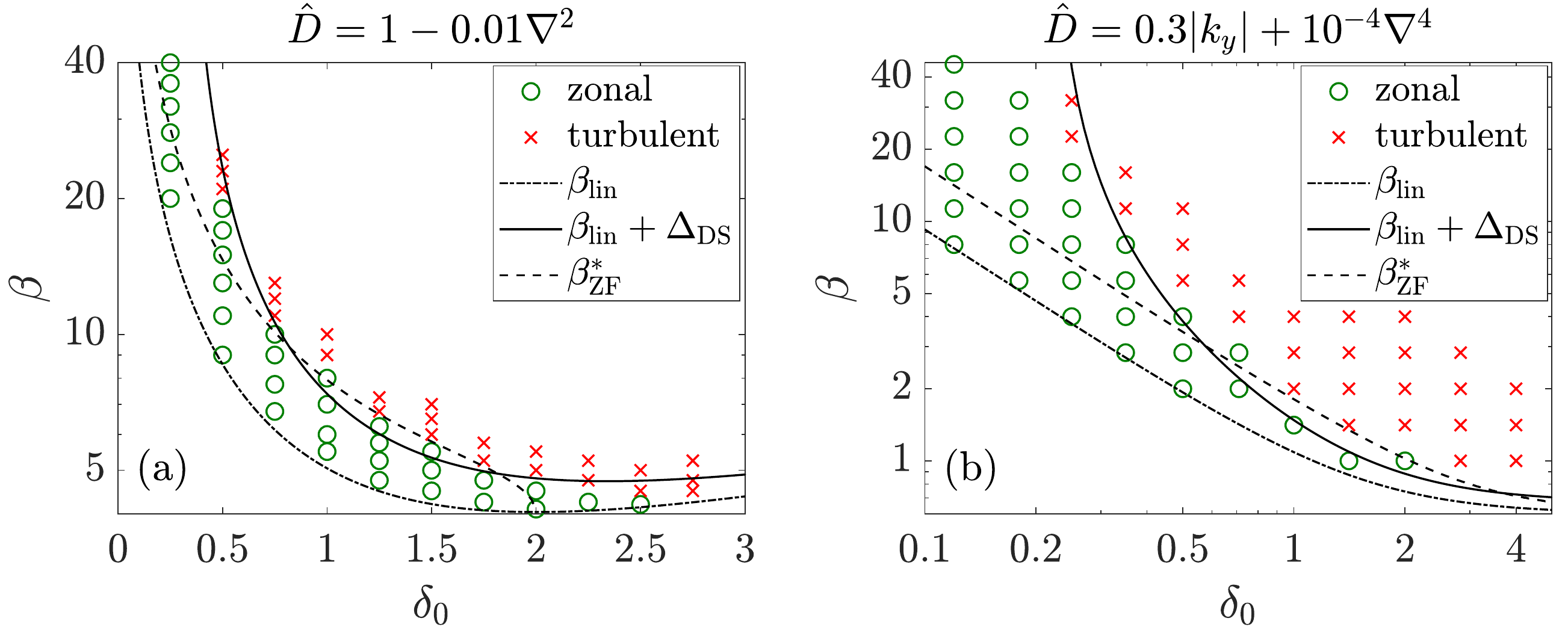}
\caption{The Dimits shift obtained by simulating the mTHE (\ref{eq:Equations_mTHE}) numerically (colored markers) versus analytic theory (black curves) for two different  choices of the damping operator: (a) $\hat{D}=1-0.01\nabla^2$ and (b) $\hat{D}=0.3|k_y|+10^{-4}\nabla^4$. Green circles indicate the Dimits regime, in which the system saturates in a state with ZFs and no turbulence. Red crosses correspond to the situation where the system remains in a turbulent state indefinitely. Dot-dashed curve: the linear threshold of the primary instability. Solid curve: our prediction of the Dimits shift, $\Delta_{\rm DS}$  (\ref{eq:DSpredction_DeltaDS}), with (a) $\varrho=0.05$ and (b) $\varrho=0.025$. Ideally, the curve $\beta_{\rm c}=\beta_{\rm lin}+\Delta_{\rm DS}$ is supposed to separate regions with green circles and with red crosses. Dashed curve (denoted $\beta_{\rm ZF}^*$): the prediction of $\beta_{\rm c}$ from \cite{St-Onge17}.}
\label{fig:DSprediction}
\end{figure}

As seen from the previous sections, the TI is nothing but the primary instability modified by nonzero ZF curvature $\mathcal{C}$. The nonzero $\mathcal{C}$ modifies the growth rate by $\Delta\gamma = \gamma_{\rm TI}(\mathcal{C}) - \gamma_{\rm TI}(0)$. We take $\gamma_{\rm TI}=\gamma_{\rm TI}^{\rm R}$  (\ref{eq:Analytic_gammaR}), since the runaway mode usually has the largest growth rate in the mTHE model. Letting $\gamma_{\rm TI}(\mathcal{C})=0$, we obtain an implicit expression for the critical value of $\beta$, denoted $\beta_{\rm c}$: 
\begin{equation}
     \beta_{\rm c}=\frac{\beta_{\rm lin}}{(1+\varrho)-\delta_0^{-1}(k_y+k_y^{-1})\sqrt{(\varrho+\varrho^2)/2}}, \quad\beta_{\rm lin}\doteq\frac{D_0[(1+k_y^2)^2+\delta_0^2k_y^2]}{\delta_0k_y^2}.
\end{equation}
Here, $\varrho\doteq\mathcal{C}/\beta_{\rm c}$, and $\beta_{\rm lin}$ is the linear threshold of the primary instability, which is obtained by letting $\gamma_{\boldsymbol{k}}=0$ (see (\ref{eq:Linear_Primary_Omegak})). Due to nonzero $\mathcal{C}$, the value of $\beta_{\rm c}$ differs from $\beta_{\rm lin}$ by a finite value $\Delta_{\rm DS}$, which represents the Dimits shift:
\begin{equation}
    \Delta_{\rm DS}=\beta_{\rm c}-\beta_{\rm lin}.\label{eq:DSpredction_DeltaDS}
\end{equation}
Note that  the chosen formula for $\gamma_{\rm TI}^{\rm R}$, (\ref{eq:Analytic_gammaR}),  is not as accurate as its counterpart (\ref{eq:GammaTI_runaway_GrowthRate}); nevertheless, we choose (\ref{eq:Analytic_gammaR}) because it does not involve the fitting parameter $\eta_{\rm R}$.
 
In section~\ref{subsec:Linear_Secondary}, we discussed the evolution of the secondary instability, where we found that the system enters a fully nonlinear stage when the ZF amplitude $u$ reaches $q_x\varphi_{\rm c}\sim\beta$ (see (\ref{eq:Linear_Secondary_phic})). Therefore,  we assume that $\mathcal{C}\sim q_x^2u$ is proportional to $\beta$; hence, $\varrho$ is assumed constant and will be treated as a fitting parameter. Then, for each value of $\delta_0$, $\Delta_{\rm DS}$ can be obtained by minimizing it over $k_y$. The results are in good agreement with  numerical simulation of the mTHE (figure~\ref{fig:DSprediction}). A similar figure can be found in figure~7 of \cite{St-Onge17}, where simulation results are compared with a different theory.

{Note that the assumption of constant $\varrho$ is not a rigorous result but only a rough approximation. In section \ref{subsec:Linear_Secondary} we showed that the ZF with $q_x=2\pi/L_x$ grows fastest as the secondary instability. However, at the fully nonlinear stage, the ZF shape is changed by additional DW--ZF interactions, and $q_x$ is no longer determined by $L_x$. As a result, the Dimits shift is insensitive to $L_x$ as long as $L_x$ is large enough. From numerical simulations, we found that the ZF shape differs from one realization to another, but in general, $q_x$ (and hence $\varrho$) is larger at smaller $\delta_0$. In fact, we also tried $\varrho=\varrho(\delta_0)$ such that $\varrho$ gets larger at smaller $\delta_0$, but the improvements in predicting the Dimits shift were not significant compared to the simpler assumption of constant $\varrho$.}

For comparison, the prediction of $\beta_{\rm c}$ made by \cite{St-Onge17} is also plotted in figure~\ref{fig:DSprediction}, where it is denoted $\beta_{\rm ZF}^*$. As a reminder, \citeauthor{St-Onge17} obtained $\beta_{\rm ZF}^*$  from a sufficient  condition  for  the  ZF  to be stable based on the 4MT approximation and considered $\beta_{\rm ZF}^*$  as a ``heuristic calculation'' of the Dimits shift. Since the 4MT method misses essential features of TI modes such as mode localization, \citeauthor{St-Onge17}'s model is less accurate than ours. Besides, the direct relation between \citeauthor{St-Onge17}'s criterion and the Dimits shift is only an assumption.  In contrast, our calculation provides an explicit formula for the Dimits shift, namely, (\ref{eq:DSpredction_DeltaDS}). Note that our (\ref{eq:DSpredction_DeltaDS}) predicts infinite $\beta_{\rm c}$ at $\delta_0= |k_y+k_y^{-1}|\sqrt{\varrho/2}$, \ie small $\delta_0$ (assuming $\varrho\ll 1$), which is in agreement with simulation results. In contrast, $\beta_{\rm ZF}^*$ is still finite in this region. Also, \citeauthor{St-Onge17}'s criterion does not have a solution at $\delta_0>|k_y+k_y^{-1}|$, suggesting zero $\Delta_{\rm DS}$; however, our theory gives nonzero $\Delta_{\rm DS}$ in this region, which is in agreement with numerical simulations.

\section{Conclusions}

In conclusion, this paper expands on our recent theory \citep{Zhu20}, where the TI and the Dimits shift were studied within reduced models of drift-wave turbulence. Here, we elaborate on a specific limit of that theory where turbulence is governed by the scalar mTHE model and the problem becomes analytically tractable. We show that assuming a sufficient scale separation between ZFs and DWs, TI modes are localized at extrema of the ZF velocity $U(x)$, where $x$ is the radial coordinate. By approximating $U(x)$ with a parabola, we analytically derive the TI growth rate, $\gamma_{\rm TI}$, using two different approaches: (i) by drawing an analogy between TI modes and quantum harmonic oscillators and (ii) by using the WME. Our theory shows that the TI is essentially a primary DW instability modified by the ZF curvature $U''$ near extream of $U$. In particular, the WME allows us to understand how the local $U''$ modifies the mode structure and reduces the TI growth rate; it also shows that the TI is \textit{not} the KHI. Then, depending on $U''$, the TI can be suppressed, in which case ZFs are strong enough to suppress turbulence (Dimits regime), or unleashed, so ZFs are unstable and turbulence develops. This understanding is different from the traditional paradigm \citep{Biglari90}, where turbulence is controlled by the flow shear $U'$.  Finally, by letting $\gamma_{\rm TI}=0$, we obtain an analytic prediction of the Dimits shift, which agrees with our numerical simulations of the mTHE.

The authors thank W. D. Dorland, N. R. Mandell, D. A. St-Onge, P. G. Ivanov, A. A. Schekochihin for helpful discussions, and the anonymous reviewers for providing numerous valuable comments. This work was supported by the US DOE through Contract No. DE-AC02-09CH11466. Digital data can also be found in DataSpace of Princeton University (\url{http://arks.princeton.edu/ark:/88435/dsp016q182p06m}).

\appendix
\section{Wigner--Moyal equation for the mTHE model}
\label{appA}

Here, we present the WME for the mTHE model following the same method that was originally used by \cite{Ruiz16} for the modified Hasegawa--Mima model. We start with the linearized DW dynamics described by (\ref{eq:Linear_Tertiary_linearDW}). Because the flow velocity $U(x,t)$ does not depend on $y$, we assume that the wave is monochromatic in $y$, namely, 
\begin{equation}
\tilde{w}=w(x,t){\rm e}^{{\rm i}k_{y}y}.
\end{equation}
Then, equation (\ref{eq:Linear_Tertiary_linearDW}) can be written
symbolically as
\begin{equation}
{\rm i}\partial_t w=\hat{H}w,\quad\hat{H}(x,\hat{k}_{x},t)=k_{y}\hat{U}+k_{y}(\beta+\hat{U}'')\hat{\bar{k}}^{-2}-{\rm i}\hat{D},\label{eq:Drifton_Shrodinger}
\end{equation}
where
\begin{equation}
    \hat{U}=U(\hat{x},t),\quad\hat{k}_x=-{\rm i}\,{\rm d}/{\rm d}x,\quad\hat{\bar{k}}^{2}=1+k_{y}^{2}+\hat{k}_x^2-{\rm i}\delta_{0}k_{y}.
\end{equation}
This can be considered as a linear Schr\"{o}dinger equation with an non-Hermitian
Hamiltonian. From here, we derive the following WME using the same phase-space formulation that is used in quantum mechanics \citep{Moyal49}:
\begin{equation}
\partial_{t}W(x,k_{x},t)=\{\!\{ \mathcal{H},W\}\!\}
+\left[\left[\Gamma,W\right]\right].\label{eq:Drifton_WignerMoyal}
\end{equation}
Here, $W$ is the Wigner function defined as
\begin{equation}
W(x,k_{x},t)\doteq\int{\rm d}s\,{\rm e}^{-{\rm i}k_x s}w^*(x-s/2,t)w(x+s/2,t)\label{eq:appA_Wigner}
\end{equation}
($^*$ denotes complex conjugate), and $\mathcal{H}$ and $\Gamma$ are the Hermitian and anti-Hermitian parts of the Hamiltonian:
\begin{subeqnarray}
    \mathcal{H}=k_y U+\Real\left(\frac{k_y\beta}{\bar{k}^{2}}\right)+\frac{k_y}{2}(U''\star\bar{k}^{-2}+\bar{k}^{*-2}\star U''),\\
    \Gamma={\rm Im}\left(\frac{k_y\beta}{\bar{k}^{2}}\right)+\frac{k_y}{2{\rm i}}(U''\star\bar{k}^{-2}-\bar{k}^{*-2}\star U'')-D_{\boldsymbol{k}},
\end{subeqnarray}
where $\bar{k}^2\doteq 1+k_y^2+k_x^2-{\rm i}\delta_0 k_y$. The symbol $\star$ is the Moyal star product:
\begin{equation}
    A\star B\doteq A\exp({\rm i}\hat{\mathcal{L}}/2) B,\quad \hat{\mathcal{L}}\doteq \frac{\overleftarrow{\partial}}{\partial x}\,\frac{\overrightarrow{\partial}}{\partial k_x}-\frac{\overleftarrow{\partial}}{\partial k_x}\,\frac{\overrightarrow{\partial}}{\partial x},
\end{equation}
where the overhead arrows in $\hat{\mathcal{L}}$ indicate the directions in which the derivatives act on, and $\lbrace\!\lbrace.,.\rbrace\!\rbrace$ and $\left[\left[.,.\right]\right]$
are the Moyal brackets:  
\begin{eqnarray}
\lbrace\!\lbrace A,B \rbrace\!\rbrace\doteq -{\rm i}(A\star B - B\star A),\quad [[A,B]]\doteq A\star B + B\star A.
\end{eqnarray}
Equation (\ref{eq:Drifton_WignerMoyal}) is mathematically equivalent
to (\ref{eq:Drifton_Shrodinger}), and the corresponding  equation for  TI eigenmodes is obtained by replacing $\partial_{t}W$ with $2\gamma_{{\rm TI}}W$. 

If we adopt the parabolic approximation of the ZF velocity, $U=U_0+\mathcal{C}x^2/2$, then $U''=\mathcal{C}$ is constant and
\begin{equation}
    \mathcal{H}=k_{y}U_{0}+\frac{1}{2}k_{y}\mathcal{C}x^{2}+k_{y}(\beta+\mathcal{C})\Real\left(\frac{1}{\bar{k}^{2}}\right),\quad\Gamma=k_{y}(\beta+\mathcal{C}){\rm Im}\left(\frac{1}{\bar{k}^{2}}\right)-D_{\boldsymbol{k}}.\label{eq:appA_Drifton_Hamiltonian}
\end{equation}
Then, the $x$-dependent part and the $k_{x}$-dependent part in $\mathcal{H}$
are separated, and $\Gamma$ is independent of $x$. This greatly simplifies the WME (\ref{eq:Drifton_WignerMoyal}),  such that it acquires the form  (\ref{eq:GammaTI_WME}), which we repeat here:
\begin{equation}
\frac{\partial W}{\partial t} = k_{y}\mathcal{C}x\,\frac{\partial W}{\partial k_{x}}-V_{{\rm g}}\,\frac{\partial W}{\partial x}+2\Gamma W+\frac{\partial Q}{\partial x}.
\end{equation}
Here, $Q$ is given by a lengthy expression,
\begin{equation}
    Q=\sum_{n=1}^{\infty}\frac{(-1)^{n+1}}{(2n+1)!\times2^{2n}}\frac{\partial^{2n+1}f}{\partial k_{x}^{2n+1}}\frac{\partial^{2n}W}{\partial x^{2n}}+\sum_{n=1}^{\infty}\frac{(-1)^{n}}{(2n)!\times2^{2n-1}}\frac{\partial^{2n}\Gamma}{\partial k_{x}^{2n}}\frac{\partial^{2n-1}W}{\partial x^{2n-1}},\label{eq:Drifton_Q}
\end{equation}
with $f(k_x)\doteq k_y U_0 +k_y(\beta+\mathcal{C})\Real(\bar{k}^{-2})$. However, $\partial_x Q$ does not contribute to the integral of (\ref{eq:Drifton_Q}) over $x$ that we are interested in. Therefore, the WME provides a transparent description of the TI under the assumption of parabolic $U$.

\section{Wave-kinetic equation and phase-space trajectories}
\label{appB}

Here, we briefly overview the derivation and the structure of drifton phase-space trajectories from the wave-kinetic equation (WKE). This discussion helps clarify the terms ``runaway mode'' and ``trapped mode'' used in the main text. It also illustrates how the TI-mode structures change with the parameter $\delta_0$. 

The WKE is an approximation of the WME in the limit when, roughly speaking, the characteristic ZF scales are much larger than the typical DW wavelength. Since a parabolic $U$ does not have a well-defined spatial scale,  we switch to the sinusoidal ZF velocity,
\begin{equation}
U=u\cos q_x x,    
\end{equation}
in which case the ZF scale is characterized by $q_x^{-1}$. For large enough ZF scale,  the WME reduces to the WKE:
\begin{equation}
\frac{\partial W}{\partial t}=\frac{\partial \mathcal{H}}{\partial x}\frac{\partial W}{\partial k_x}-\frac{\partial \mathcal{H}}{\partial k_x}\frac{\partial W}{\partial x}+2\Gamma W,\label{eq:Drifton_iWKE}
\end{equation}
where 
\begin{equation}
\mathcal{H}= k_{y}\left[1-\Real\left(\frac{q_x^2}{\bar{k}^{2}}\right)\right]u\cos q_x x+k_{y}\beta\Real\left(\frac{1}{\bar{k}^{2}}\right),\label{eq:appB_Drifton_Hamiltonian}
\end{equation}
while $\Gamma$ is not important for the following discussions. The form of the WKE (\ref{eq:Drifton_iWKE}) indicates that $W$ can be considered as the distribution function of DW quanta, or driftons,  in the $(x,k_{x})$ phase space. The driftons trajectories are governed by Hamilton's equations,
\begin{equation}
    \frac{{\rm d}x}{{\rm d}t}=\frac{\partial\mathcal{H}}{\partial k_x},\quad  \frac{{\rm d}k_x}{{\rm d}t}=-\frac{\partial\mathcal{H}}{\partial x},
\end{equation}
where $\mathcal{H}$ serves as the Hamiltonian.  However, unlike true particles, driftons are not conserved. Instead, $\Gamma$ determines the rate at which $W$ evolves along the ray trajectories. 

If ZFs are stationary, as is the case for our calculation of the TI, then $\mathcal{H}$
is independent of time and driftons move along curves that satisfy $\mathcal{H}(x, k_x) = \mathcal{E}$, where $\mathcal{E}$ is a constant.  In \cite{Zhu18b}, we systematically studied these trajectories for the modified Hasegawa--Mima system ($\hat{\delta}=0$), and
three types of trajectories have been identified, which we called passing, trapped, and
runaway trajectories. Although the mTHE has nonzero $\hat{\delta}$, it corresponds to similar drifton dynamics unless $\hat{\delta}$ is too large.  Note that $\mathcal{H}$ depends on $\Real(1/\bar{k}^{2})$, which is
\begin{equation}
\Real\left(\frac{1}{\bar{k}^{2}}\right)=\frac{1+k_{x}^{2}+k_{y}^{2}}{(1+k_{x}^{2}+k_{y}^{2})^{2}+\delta_{0}^{2}k_{y}^{2}}.
\end{equation}
Therefore, $\Real(1/\bar{k}^{2})$ is a monotonically decreasing
function of $k_{x}^{2}$ if $\delta_{0}^{2}k_{y}^{2}<(1+k_{y}^{2})^{2}$,
\ie when $\delta_{0}<|k_{y}+k_{y}^{-1}|$.
However, $\Real(1/\bar{k}^{2})$ has a maximum at nonzero $k_{x}^{2}$
if $\delta_{0}>|k_{y}+k_{y}^{-1}|\geq2$. 
In the following, we discuss the two situations separately.

\begin{figure}
\includegraphics[width=1\columnwidth]{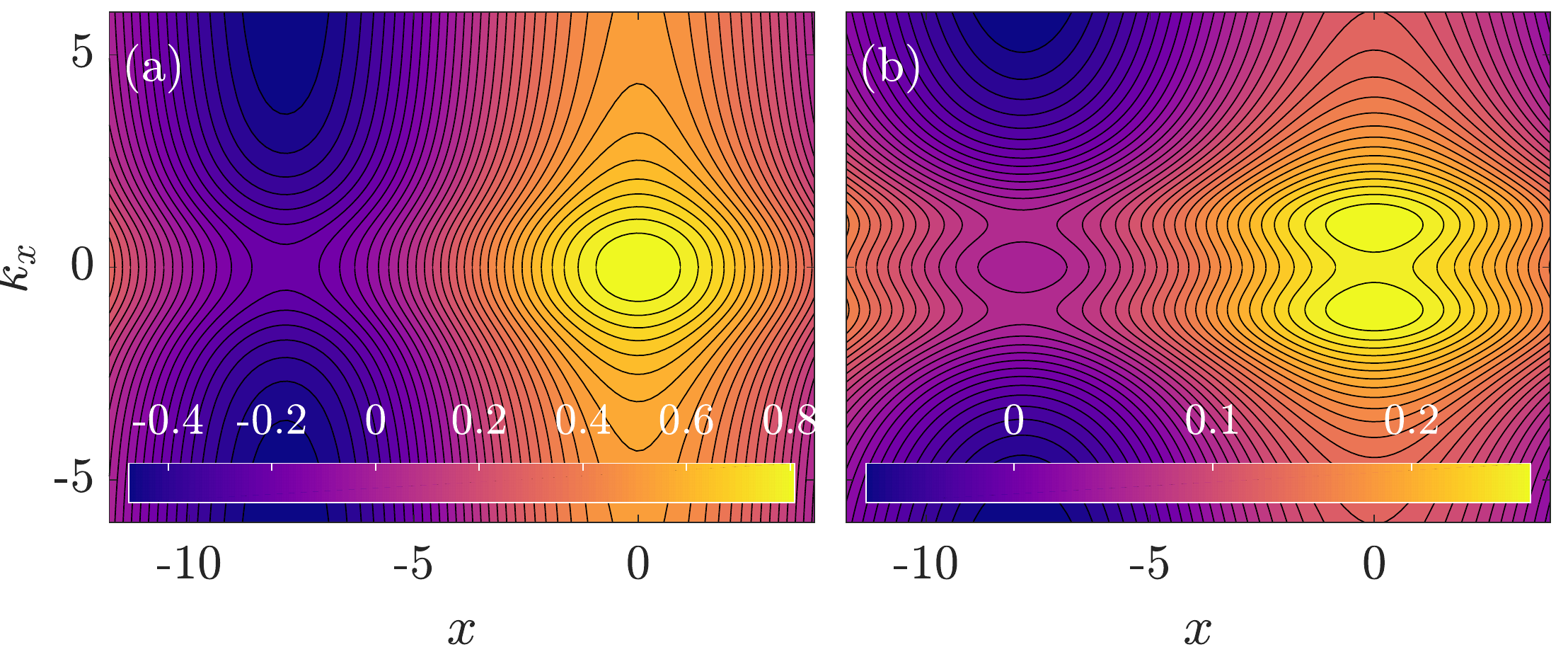}\caption{Contour plots of the drifton Hamiltonian $\mathcal{H}$ (\ref{eq:appB_Drifton_Hamiltonian}) at (a) small $\delta_{0}$ and (b) large $\delta_{0}$; the color marks the corresponding value of $\mathcal{H}$. The parameters are $\beta=k_{y}=1$, $q_x=0.4$; also, (a) 
$\delta_{0}=1.5$ and $u=0.5$, and (b) 
$\delta_{0}=3$ and $u=0.1$. At small $\delta_{0}$, trapped trajectories are found near the ZF maximum $x=0$ and runaway
trajectories are found near the ZF minimum $x=\upi/q_{x}$. At large $\delta_{0}$, two separate trapped islands form at $x=0$ and trapped
trajectory replace runaway trajectories at $x=\upi/q_x$.\label{fig:THtraj}}
\end{figure}

First, consider $\delta_{0}<|k_{y}+k_{y}^{-1}|$. Then, letting
$\mathcal{H}=\mathcal{E}$ leads to
\begin{equation}
k_{x}^{2}(x,\mathcal{E})=(1+k_{y}^{2})\frac{\mathcal{H}^{0}(x,\mathcal{E})-\mathcal{E}}{\mathcal{E}-\mathcal{H}^{\infty}(x)},\label{eq:Drifton_px(x)}
\end{equation}
where
\begin{eqnarray}
&\mathcal{H}^{\infty}(x)\doteq k_{y}u\cos q_{x}x,
\\
&\mathcal{H}^{0}(x,\mathcal{E})=k_{y}u\cos q_{x}x+\frac{k_{y}}{1+k_y^2}(\beta-q_{x}^{2}u\cos q_{x}x)
\left(1-\frac{\lambda}{2}\right),
\end{eqnarray}
and
\begin{equation}
\lambda=\lambda_{\pm}(x,\mathcal{E})\doteq1\pm\sqrt{1-\frac{4\delta_{0}^{2}(\mathcal{H}^{\infty}-\mathcal{E})^2}{(\beta-q_{x}^{2}u\cos q_{x}x)^{2}}}.
\end{equation}
This shows that at given $x$, there are two solutions for $k_{x}^{2}$
depending on whether $\lambda=\lambda_{+}$ or $\lambda=\lambda_{-}$.
However, it turns out that $\lambda=\lambda_{+}$ corresponds to negative $k_{x}^{2}$
and hence can be ignored, which is consistent with the fact that $\mathcal{H}$
is a monotonic function of $k_{x}^{2}$ at small $\delta_{0}$. Therefore,
only $\lambda=\lambda_{-}$ is possible, and one could use (\ref{eq:Drifton_px(x)})
to identify passing, trapped, and runaway trajectories as in \cite{Zhu18b}. At very small $u$, ZFs do not matter, so all trajectories are passing. However, when $u$ exceeds a certain critical amplitude $u_{\rm c}$, passing trajectories disappear, which indicates that DWs are strongly affected by ZFs in this case. The critical ZF amplitude is obtained by letting
\begin{equation}
\max\mathcal{H}^{\infty}=\min\mathcal{H}^{0}.
\end{equation}
This leads to  an implicit expression of $u_{\rm c, 1}$:
\begin{equation}
u_{{\rm c}}=\frac{\beta}{2(1+k_{y}^{2})-q_{x}^{2}}\left[1-\frac{\lambda_{0}}{2}\left(1+\frac{q_{x}^{2}u_{\rm c}}{\beta}\right)\right],\label{eq:appB_U_critical}
\end{equation}
where
\begin{equation}
\lambda_{0}\doteq\lambda_{-}(x=0,\mathcal{E}=k_{y}u_{\rm c}).
\end{equation}
Therefore, $u_{{\rm c}}$ is smaller than that in the modified
Haseagawa--Mima system, where $\lambda_{0}=0$ \citep{Zhu18b}.  Phase-space trajectories at $u>u_{\rm c}$ are shown in figure~\ref{fig:THtraj}(a).

At $\delta_{0}>|k_{y}+k_{y}^{-1}|\geq2$, $\lambda=\lambda_{-}$
still gives passing and runaway trajectories as before. However, because
$\mathcal{H}$ becomes non-monotonic with respect to $k_{x}^{2}$,
the other solution $\lambda=\lambda_{+}$ can also give positive $k_{x}^{2}$
for some values of $\mathcal{E}$. As a result, runaway trajectories
are replaced with trapped trajectories near the ZF minimum, and two separate
trapped islands are formed near the ZF maximum.   The corresponding phase-space trajectories are shown in figure~\ref{fig:THtraj}(b).

\bibliographystyle{jpp}

\bibliography{Ref}

\end{document}